\documentclass[preprint2]{aastex}
\usepackage{natbib}
\lefthead{Hayashi et al}
\righthead{Inner $\Lambda$CDM halo structure and LSB rotation curves}

\usepackage{epsfig}

\newcommand{\etal}{{et al.~}}

\def\lsim{\mathrel{\hbox{\rlap{\hbox{\lower4pt\hbox{$\sim$}}}\hbox{$<$}}}}
\def\gsim{\mathrel{\hbox{\rlap{\hbox{\lower4pt\hbox{$\sim$}}}\hbox{$>$}}}}

\def\nhalos{seven }

\def\LCDM{$\Lambda$CDM }

\newcommand{\kms}{\>{\rm km}\,{\rm s}^{-1}}

\newcommand{\kpch}{\>h^{-1} {\rm kpc}}
\newcommand{\kpc}{\>{\rm kpc}}
\newcommand{\Mpch}{\>h^{-1} {\rm Mpc}}

\newcommand{\Msolh}{\>h^{-1} M_{\odot}}

\begin{document}

\title
{The Inner Structure of $\Lambda$CDM Halos II: Halo Mass Profiles and LSB
Rotation Curves}
\author{Eric Hayashi\altaffilmark{1}, Julio
F.~Navarro\altaffilmark{1,6}, 
Chris~Power\altaffilmark{2},
Adrian~R. Jenkins\altaffilmark{2},\\
Carlos S.~Frenk\altaffilmark{2},
Simon D.~M.~White\altaffilmark{3},
Volker~Springel\altaffilmark{3},\\
Joachim~Stadel\altaffilmark{4} and Thomas~R. Quinn\altaffilmark{5}}

\affil{$^1$Department of Physics and Astronomy, University of Victoria, 
 Victoria, BC, V8P 1A1, Canada}

\affil{$^2$ Institute for Computational Cosmology, Department of Physics,
  University of Durham, South Road, Durham, DH1 3LE}

\affil{$^3$ Max-Planck Institute for Astrophysics, Garching, Munich, D-85740, Germany}

\affil{$^4$ Institute for Theoretical Physics, University of Zurich, Zurich  CH-8057, Switzerland}

\affil{$^5$ Department of Astronomy, University of Washington, Seattle, WA
98195, USA.}


\altaffiltext{6}{
Fellow of the Canadian Institute for Advanced Research and of the J.S.Guggenheim
Memorial Foundation}



\begin{abstract}
We use a set of high-resolution cosmological N-body simulations to investigate
the inner mass profile of galaxy-sized cold dark matter (CDM) halos. These
simulations extend the thorough numerical convergence study presented in Paper I
of this series \citep{POWER03}, and demonstrate that the mass profile of CDM
halos can be robustly estimated beyond a minimum converged radius of order
$r_{\rm conv} \sim 1~\kpch$ in our highest resolution runs. The density profiles
of simulated halos become progressively shallow from the virial radius inwards,
and show no sign of approaching a well-defined power-law behaviour near the
centre. At $r_{\rm conv}$, the logarithmic slope of the density profile is
steeper than the asymptotic $\rho\propto r^{-1}$ expected from the formula
proposed by Navarro, Frenk, and White (1996), but significantly shallower than
the steeply divergent $\rho\propto r^{-1.5}$ cusp proposed by \cite{MOORE99}.
We perform a direct comparison of the spherically-averaged dark matter circular
velocity ($V_c$) profiles with rotation curves of low surface brightness (LSB)
galaxies from the samples of \cite{DEBLOK01}, \cite{DEBLOK02}, and
\cite{SWATERS03}.  Most (about two-thirds) LSB galaxies in this dataset are
roughly consistent with CDM halo $V_c$ profiles.  However, about one third of
LSBs in these samples feature a sharp transition between the rising and flat
part of the rotation curve that is not seen in the $V_c$ profiles of CDM
halos. This discrepancy has been interpreted as excluding the presence of cusps,
but we argue that it might simply reflect the difference between circular
velocity and gas rotation speed likely to arise in gaseous disks embedded within
realistic, triaxial CDM halos.
\end{abstract}


\section{Introduction}
\label{sec:intro}

The structure of dark matter halos and its relation to the cosmological context
of their formation has been studied extensively over the past few decades. Early
analytic calculations focused on the scale free nature of the gravitational
accretion process and suggested that halo density profiles might be simple power
laws \citep{GUNN72, FILLMORE84, HOFFMAN85, WHITE92}. Cosmological N-body
simulations, however, failed to confirm these analytic expectations. Although
power-laws with slopes close to those motivated by the theory were able to
describe some parts of the halo density profiles, even early simulations found
that significant deviations from a single power-law behaviour were present in
most cases \citep{FRENK85, FRENK88, QUINN86, DUB91, CRONE94}. Further simulation
work, indeed, concluded that power-law fits were inappropriate, and that,
properly scaled, dark halos spanning a wide range in mass and size are well fit
by a ``universal'' density profile \citep[hereafter NFW]{NFW95, NFW96, NFW97}:
\begin{equation}
\label{eq:nfw}
\rho_{\rm NFW}(r) =  \frac{\rho_s}{(r/r_s)(1 + r/r_s)^2}.
\end{equation}

One characteristic of this fitting formula is that the logarithmic slope,
$\beta(r)=-d\log\rho/d\log r=(1+3y)/(1+y)$ (where $y=r/r_s$ is the radius in
units of a characteristic scale radius, $r_s$), increases monotonically from the
centre outwards. The density profile steepens with increasing radius; it is
shallower than isothermal inside $r_s$, and steeper than isothermal for
$r>r_s$. Another important feature illustrated by this fitting formula is that
the profiles are ``cuspy''($\beta_0=\beta(r=0)>0$): the dark matter density (but
not the potential) diverges formally at the centre.

Subsequent work has generally confirmed these trends, but has also highlighted
potentially important deviations from the NFW fitting formula. In particular,
\cite{FUK97,FUK01}, as well as Moore and collaborators
\citep{MOORE98,MOORE99,GHIGNA00}, have reported that NFW fits to their simulated
halos (which had much higher mass and spatial resolution than the original NFW
work) underestimate the dark matter density in the innermost regions
($r<r_s$). These authors proposed that the disagreement was indicative of inner
density ``cusps'' steeper than the NFW profile and advocated a simple
modification to the NFW formula so that $\beta_0=1.5$ (rather than $1.0$).

The actual value of the asymptotic slope, $\beta_0$, is still being hotly
debated in the literature \citep{JING95,KLY01,TAYLOR01,NAVARRO03,POWER03,FUK03},
but there is general consensus that CDM halos are indeed ``cuspy''. This has
been recognized as an important result, since the rotation curves of many disk
galaxies, and in particular of low surface brightness (LSB) systems, appear to
indicate the presence of an extended region of constant dark matter density: a
dark matter ``core''
\citep{FLORES94,MOORE94,BURKERT95,BLAIS01,DEBLOK01A,DEBLOK01}.

Unfortunately, rotation curve constraints are strongest just where numerical
simulations are least reliable.  Resolving CDM halos down to the kpc scales
probed by the innermost points of observed rotation curves requires extremely
high mass and force resolutions, as well as careful integration of particle
orbits in the central, high density regions of halos. This poses a significant
computational challenge that has been met in very few of the simulations
published to date.

This difficulty has meant that rotation curves have usually been compared with
extrapolations of the simulation data that rely heavily on the applicability and
accuracy of fitting formulae such as the NFW profile to regions that may be
severely compromised by numerical artifact. This practice does not
allow for halo-to-halo variations, or for temporary departures from the
``average'' profile to be properly taken into account when trying to model the
observational datasets.

Finally, the theoretical debate on the asymptotic central slope of the dark
matter density profile, $\beta_0$, has led at times to unwarranted emphasis on
the very inner region of the rotation curve datasets, rather than on an proper
appraisal of the data over its full radial extent. For example,
\cite{DEBLOK01A,DEBLOK01} attempt to derive constraints on $\beta_0$ from the
innermost few points of their rotation curves, and conclude that $\beta_0\sim 0$
for most galaxies in their sample. However, their analysis focuses on the
regions most severely affected by non-circular motions, seeing, misalignments
and slit offsets, and other effects that limit the accuracy of circular velocity
estimates based on long-slit spectra. It is perhaps not surprising, then, that
other studies have disputed the conclusiveness of these findings.

Indeed, an independent analysis of data of similar quality by \cite{SWATERS03}
\citep[see also][]{VDB00}  concludes that ``cuspy'' dark matter halos with
$\beta_0 \lsim 1$ are actually consistent with their data. This disagreement
is compounded by the results of the latest cosmological N-body simulations
\citep[hereafter P03]{POWER03}, which find scant evidence for a well defined
value of $\beta_0$ in simulated CDM halos. Given these difficulties, focusing
the theoretical or observational analysis on $\beta_0$ seems unwise. 

In this paper, we improve upon previous work by comparing directly the actual
results of the simulations with the full radial extent of the rotation curves of
LSB galaxies. We present results from a set of \nhalos galaxy-sized dark matter
halos, each of which has been carefully simulated at various resolution levels
in order to ascertain the numerical convergence of our results. This allows us
to test rigorously the P03 convergence criteria, as well as to assess the
cusp-core discrepancy through direct comparison between observation and
simulations. A companion paper \citep{NAVPREP03} addresses the issue of
universality of CDM halo structure using simulations that span a wide range of
scales, from dwarf galaxies to galaxy clusters.

The outline of this paper is as follows.  In \S\ref{sec:simulations} we
introduce our set of simulations and summarize briefly our numerical
methods. The seven galaxy-sized halos that form the core of our sample have been
simulated at various resolutions, and we use them in \S\ref{sec:convcrit} to
investigate the robustness of the P03 numerical convergence criteria. The
density profiles of these halos are presented and compared with previous work in
\S\ref{sec:profshape}.  In \S\ref{sec:rotcurve} we compare the halo $V_c$
profiles with the LSB rotation curve datasets of \cite{DEBLOK01},
\cite{DEBLOK02}, and \cite{SWATERS03}.  Our main conclusions and plans for
future work are summarized in \S\ref{sec:concl}.

\section{The Numerical Simulations}
\label{sec:simulations}

We have focused our analysis on \nhalos galaxy-sized dark matter halos selected
at random from two different cosmological N-body simulations of large periodic
boxes with comoving size $L_{\rm box}=32.5 \Mpch$ and $35.325 \Mpch$,
respectively.  Each of these ``parent'' simulations has $N_{\rm box} = 128^3$
particles, and adopts the currently favoured flat, low-density ``concordance''
\LCDM world model, with $\Omega_0=0.3$, $\Omega_\Lambda=0.7$, and either
$h=0.65$ (runs labelled G1, G2 and G3) or $h=0.7$ (G4, G5, G6, and G7, see
Table~\ref{tab:sims}). The power spectrum in both simulations is normalized so
that the linear rms amplitude of fluctuations on spheres of radius $8 \Mpch$ is
$\sigma_8=0.9$.


All halos (G1 to G7) have been re-simulated at three or four different mass
resolution levels; each level increases the number of particles in the halo by a
factor of $8$, so that the mass per particle has been varied by a factor $512$
in runs G1-G3, and by a factor $64$ in runs G4-G7 (see
Table~\ref{tab:sims}). All of these runs focus numerical resources on the
Lagrangian region from where each system draws its mass, whilst approximating
the tidal field of the whole box by combining distant particles into groups of
particles whose mass increases with distance from the halo. This resimulation
technique follows closely that described in detail in P03 and in
\cite{NAVPREP03}, where the reader is referred to for full details. For
completeness, we present here a brief account of the procedure.

Halos selected for resimulation are identified at $z=0$ from the full list of
halos with circular velocities in the range ($150$, $250$) km s$^{-1}$ in the
parent simulations.  All particles within a sphere of radius
$3~r_{200}${\footnote{We define the ``virial radius'', $r_{200}$, as the radius
of a sphere of mean density $200$ times the critical value for closure,
$\rho_{\rm crit}=3\,H^2/8\pi G$, where $H$ is Hubble's constant. We parameterize
the present value of Hubble's constant $H$ by $H_0=100\,h$ km s$^{-1}$
kpc$^{-1}$}} centred in each halo are then traced back to the initial redshift
configuration ($z_i=49$).

The region defined by these particles is typically fully contained within a box
of size $L_{\rm sbox}\simeq 5 \Mpch$, which is loaded with $N_{\rm sbox}=32^3$,
$64^3$, $128^3$, or $256^3$ particles. Particles in this new high-resolution
region are perturbed with the same waves as in the parent simulation as well as
with additional smaller scale waves up to the Nyquist frequency of the high
resolution particle grid.  Particles which do not end up within $3~r_{200}$ of
the selected halo at $z=0$ are replaced by lower resolution particles which
replicate the tidal field acting on the high resolution particles.  This
resampling includes some particles within the boundaries of the high resolution
box, and therefore the high resolution region defines an asymmetrical
``amoeba-shaped'' three-dimensional volume surrounded by tidal particles whose
mass increases with distance from this region.

A summary of the numerical parameters and halo properties is given in
Table~\ref{tab:sims}. This table also includes reference to $12$ further runs,
four of them corresponding to dwarf galaxy halos and eight of them to galaxy
cluster-sized halos. These systems have been simulated only at the highest
resolution ($N_{\rm sbox}=256^3$), and therefore are not included in our
convergence analysis. These runs are discussed in detail in a companion paper
\citep{NAVPREP03}.

Some simulations were performed with a fixed number of timesteps for all
particles using Stadel and Quinn's parallel N-body code {\tt PKDGRAV}
\citep{STADEL01}, while others used the N-body code {\tt GADGET}
\citep{SPRINGEL01}. The {\tt GADGET} runs allowed for individual timesteps for
each particle assigned using either the {\tt RhoSgAcc} or {\tt EpsAcc} criterion
(see P03 for full details). The halo labelled G1 in this paper is the same one
selected for the numerical convergence study presented in P03.  Although {\tt
PKDGRAV} also has individual timestepping capabilities, we have chosen not to
take advantage of these for the simulations presented in this paper.  We note
that P03 finds only a modest computational gain due to multi-stepping schemes
provided that the softening parameter is properly chosen.

The softening parameter (fixed in comoving coordinates) for each simulation
(with the exception of G1/$256^3$, see P03) was chosen to match the ``optimal''
softening suggested by P03:
\begin{equation}
\epsilon_{\rm opt} = \frac{4~r_{200}}{N_{200}^{1/2}},
\label{eq:optsoft}
\end{equation}
where $N_{200}$ is the number of particles within $r_{200}$ at $z=0$. This
softening choice minimizes the number of timesteps required for convergence
results by minimizing discreteness effects in the force calculations whilst
ensuring adequate force resolution.

At $z=0$, the mass within the virial radius, $M_{200}$, of our galaxy-sized halos
ranges from $\sim 10^{12} \Msolh$ to $\sim 3 \times 10^{12} \Msolh$,
corresponding to circular velocities, $V_{200} = (G M_{200}/r_{200})^{1/2}$, in
the range $160 \kms$ to $230 \kms$.

\section{Numerical Convergence}
\label{sec:convcrit}

\subsection{Criteria}

P03 propose three different conditions that should be satisfied in order to
ensure convergence in the circular velocity profile.  According to these
criteria, convergence to better than $10\%$ in the spherically-averaged circular
velocity, $V_c(r)$, is achieved at radii which satisfy the following conditions:

\begin{enumerate}

\item The local orbital timescale $t_{\rm circ}(r)$ is much greater than the size of
the timestep $\Delta t$:
\begin{equation}
\frac{t_{\rm circ}(r)}{t_{\rm circ}(r_{200})} \gsim 15 \left (\frac{\Delta t}{t_0}\right)^{5/6} 
\label{eq:tcirc}
\end{equation}

where $t_0$ denotes the age of the universe, which is by definition of the order
of the circular orbit timescale at the virial radius, $t_{\rm circ}(r_{200})$.

\item Accelerations do not exceed a characteristic acceleration, $a_{\epsilon}$,
determined by $V_{200}$ and the softening length $\epsilon$:
\begin{equation}
a(r) =\frac{V_c^2(r)}{r} \lsim a_\epsilon = 0.5~\frac{V_{200}^2}{\epsilon}
\label{eq:acceps}
\end{equation}

where $a(r)$ is the mean radial acceleration experienced by particles at a
distance $r$ from the centre of the system, $a(r)=G M(r)/r^2 = V_c^2(r)/r$.

\item 
Enough particles are enclosed such that the local collisional relaxation
timescale $t_{\rm relax}(r)$ is longer than the age of the universe{\footnote{We
adopt a slightly more conservative criterion than P03, who require $t_{\rm
relax} \gsim 0.6~t_{\rm circ}(r_{200})$.}}:
\begin{equation}
\frac{t_{\rm relax} (r)}{t_{\rm circ}(r_{200})} = \frac{\sqrt{200}}{8} \frac{N(r)}{\ln
  {N(r)}} \left (\frac{\overline{\rho}(r)}{\rho_{\rm crit}}\right)^{-1/2} \gsim 1
\label{eq:trelax}
\end{equation}
where $N(r)$ is the number of particles and $\overline{\rho}(r)$ is the mean
density within radius $r$.

\end{enumerate}

For ``optimal'' choices of the softening and timestep, as well as for the
typical number of particles in our runs, we find that criterion (3) above is the
strictest one. 

The number of high-resolution particles thus effectively defines the
``predicted'' converged radius, $r_{\rm conv}$, beyond which, according to P03,
circular velocities should be accurate to better than $10\%$. We emphasize that
this accuracy criterion applies to the cumulative mass profile; convergence in
properties such as local density estimates, $\rho(r)$, typically extends to
radii significantly smaller than $r_{\rm conv}$.

\subsection{Validating the Convergence Criteria}
\label{sec:testconvcrit}

\begin{figure}
\plotone{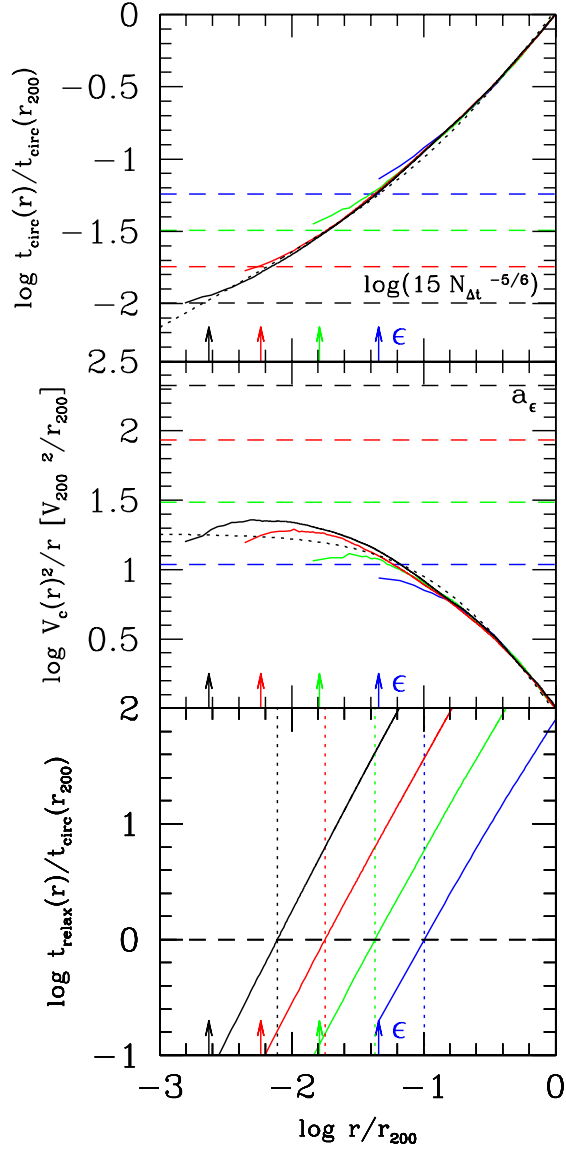}
\figcaption{
The structure of halo G3 at different mass, time, and spatial resolutions.
The value of the softening parameter $\epsilon$ is indicated by arrows in the
three panels. The number of timesteps and particles are listed in
Table~\ref{tab:sims}.  Runs with $32^3$, $64^3$, $128^3$ and $256^3$
high-resolution particles are shown in blue, green, red, and black,
respectively. Dotted curves show an NFW profile with concentration $c=5.3$ and
$r_{200}=143.4~\kpch$. {\it Upper panel:} Circular orbital timescale $t_{\rm
circ}$ versus radius; radii at which the circular orbital timescale is less than
$15~N_{\Delta t}^{-5/6}$, indicated by the dashed lines for each simulation, are
unresolved due to insufficient time resolution.  {\it Middle panel:} Mean radial
acceleration profile $V_c(r)^2/r$; untrustworthy radii are those corresponding
to accelerations greater than the limiting acceleration imposed by the softening
$a_\epsilon \simeq 0.5~V_{200}^2/\epsilon$, shown by the dashed lines.  {\it
Lower panel:} Collisional relaxation time $t_{\rm relax}$ versus radius;
convergence requires $t_{\rm relax} \gsim t_{\rm circ}(r_{200})$.  Vertical
dotted lines indicate the radius, $r_{\rm conv}$, beyond which this condition
(the strictest of the three) is satisfied.
\label{fig:allcrit}}
\end{figure}

We assess the validity of the convergence criteria listed above by comparing the
mass profile of the highest resolution run corresponding to each halo with those
obtained at lower resolution. Figure~\ref{fig:allcrit} illustrates the
procedure. From top to bottom, the three panels in this figure show, as a
function of radius, the circular orbit timescale, the mean radial acceleration,
and the relaxation timescale, respectively, for the four runs corresponding to
halo G3.  The small arrows at the bottom of each panel indicate the choice of
gravitational softening for each run.  The dotted curves in the top and middle
panel show the best fit NFW profile to the converged region of the highest
resolution $N_{\rm sbox}=256^3$ run.

The ``converged radius'' corresponding to each criterion is determined by the
intersection of the horizontal dashed lines in each panel with the ``true''
profile, which we shall take to be that of the highest resolution run (shown in
solid black in Figure~\ref{fig:allcrit}). Clearly, the strictest criterion is
that imposed by the relaxation timescale (the dotted vertical lines in the lower
panel show the converged radius corresponding to this criterion).  This
suggests, for example, that the lowest-resolution G3 run (with $N_{\rm
sbox}=32^3$, shown in solid blue), should start to deviate from the converged
profiles roughly at $r\sim 0.1~r_{200}$. Indeed, this appears to be the radius
at which this profile starts to ``peel off'' from the highest resolution one, as
shown in the top two panels of Figure~\ref{fig:allcrit}.  Increasing the number
of high resolution particles by a factor of eight typically brings the converged
radius inwards by a factor of $\sim 2.4$. For the medium-resolution run ($N_{\rm
sbox}=64^3$, shown in solid green), $r_{\rm conv}$ is predicted to be $\sim
0.04~r_{200}$, which, again, coincides well with the radius inside which
departures from the converged profile are apparent. Similarly, $r_{\rm conv}\sim
0.017~r_{200}$ for the high-resolution ($N_{\rm sbox}=128^3$) run (shown in
red).

The density and circular velocity profiles corresponding to the four G3 runs are
shown in Figure~\ref{fig:allrhonprt}. Panels on the left show the profiles down
to the radius that contains $50$ particles, whereas those on the right
show the profiles restricted to $r \gsim r_{\rm
conv}$. Figure~\ref{fig:allrhonprt} illustrates two important results alluded to
above: (i) both $\rho(r)$ and $V_c(r)$ converge well at $r\gsim r_{\rm conv}$,
and (ii) convergence in $\rho(r)$ extends to radii smaller than $r_{\rm
conv}$. Indeed, the top-left panel shows that our choice of $r_{\rm conv}$ is
rather conservative when applied to the density profile. Typically, densities
are estimated to better than $10\%$ down to $r\sim 0.6\, r_{\rm conv}$.

\begin{figure}
\plotone{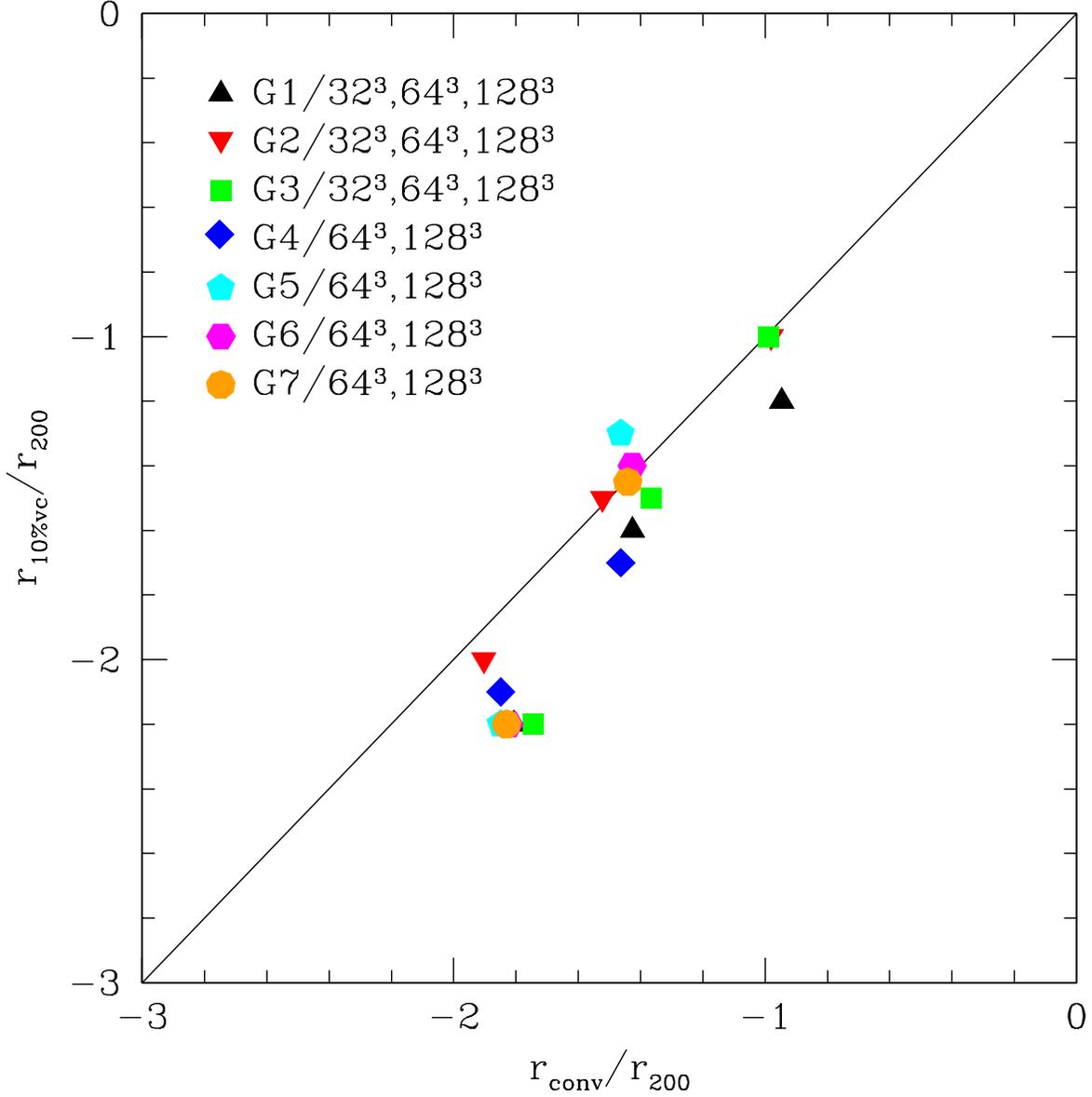}
\figcaption{
Radius where the circular velocity profile of lower resolutions runs starts to
deviate from that of the highest resolution run by more than $10\%$,
$r_{10\%vc}$, plotted against the minimum converged radius predicted by the P03
convergence criteria, $r_{\rm conv}$.  In all cases $r_{\rm conv} \leq
r_{10\%vc}$, validating the P03 criteria as conservative estimators of the
converged region.\label{fig:rminrdev}}
\end{figure}

How general are these results?  Figure~\ref{fig:rminrdev} compares the minimum
``converged'' radius predicted by the P03 criteria , $r_{\rm conv}$, with
$r_{10\%vc}$, the {\it actual} radius where circular velocities in the lower
resolution runs deviate from convergence by more than $10\%$. In essentially all
cases, $r_{\rm conv} \lsim r_{10\%vc}$, indicating that the P03 criteria are
appropriate, albeit at times somewhat conservative. We list our $r_{\rm conv}$
estimates for all runs in Table~\ref{tab:sims}.

\section{Halo Structure and Fitting Formulae}
\label{sec:profshape}

\begin{figure}
\plotone{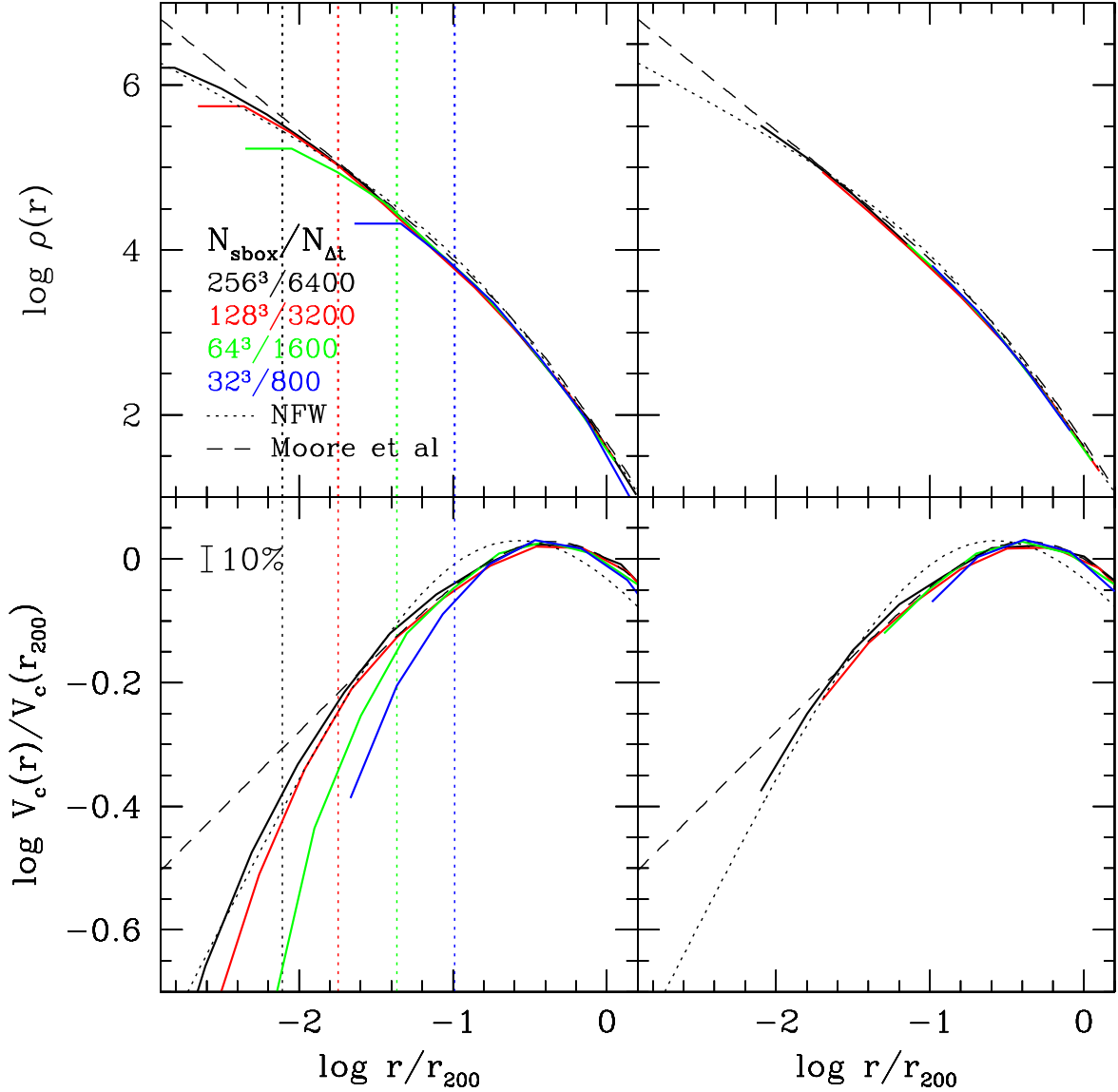}
\figcaption{
{\it Upper left panel:} Density profiles of halo G3 at four different levels of
mass resolution, plotted down to radii containing $50$ particles.  Dashed and
dotted curves show best fit Moore \etal and NFW profiles, respectively (see text
for fitting details). Vertical dotted lines indicate the minimum converged
radius, $r_{\rm conv}$, for each run. {\it Upper right panel:} Same density
profiles plotted only for converged radii.  The discrepancy between lower
resolution runs and the highest resolution simulation at small radii is no
longer apparent when only reliably resolved radii are considered.  Lower panels
are as the upper ones, but for the circular velocity profiles.
\label{fig:allrhonprt}}
\end{figure}

The dotted curves in Figure~\ref{fig:allrhonprt} show the best NFW fits to the
density and circular velocity profile of the highest resolution run. The dashed
lines correspond to the best fit adopting the modification to the NFW profile
advocated by \cite{MOORE99},
\begin{equation}
\rho_{\rm Moore}(r) = \frac{\rho_M}{(r/r_M)^{1.5} (1 + (r/r_M)^{1.5})}.
\label{eq:mooreprof}
\end {equation}
These fits are obtained by straightforward $\chi^2$ minimization in two
parameters, $r_s$ or $r_M$, and the characteristic density $\rho_s$ or
$\rho_M$. The profiles are calculated in bins of equal width in $\log r$, and
the fits are performed over the radial range $r_{\rm conv} < r < r_{200}$.
Equal weights are assigned to each radial bins because the statistical (Poisson)
uncertainty in the determination of the mass within each bin is negligible (each
bin contains thousands of particles) so uncertainties are completely dominated
by systematic errors whose radial dependence is difficult to assess
quantitatively.

The best fits to $\rho(r)$ and $V_c(r)$ shown in Figure~\ref{fig:allrhonprt} are
obtained independently from each other.  Values of the concentration parameter,
$c_{\rm NFW}=r_{200}/r_s$, for the best fit NFW profiles are 6.4 and 5.3 for
fits to the density and circular velocity profile, respectively; the Moore et al
concentrations, $c_{\rm Moore}=r_{200}/r_M$, are 3.0 and 2.9 for the best fits to
$\rho(r)$ and $V_c(r)$, respectively.  Over the converged region, $r\gsim r_{\rm
conv}$, both the NFW and Moore et al profiles appear to reproduce reasonably
well the numerical simulation results. Indeed, no profile in the G3 runs
deviates by more than $10\%$ in $V_c$ or $30\%$ in $\rho(r)$ from the best fits
obtained with either eq.~\ref{eq:nfw} or eq.~\ref{eq:mooreprof}. More
substantial differences are expected only well inside $r_{\rm conv}$, but these
regions are not reliably probed by the simulations.

This suggests that either the NFW or Moore et al profile may be used to describe
the structure of $\Lambda$CDM halos outside $\sim 1\%$ of the virial radius, but
also implies that one should be extremely wary of extrapolations inside this
radius. One intriguing feature of Figure~\ref{fig:allrhonprt} is that the Moore
et al formula appears to fit the G3 density profiles better than NFW but that
$V_c$ profiles are somewhat better approximated by NFW (see also P03). This
suggests that neither formula captures fully and accurately the radial
dependence of the structure of $\Lambda$CDM halos.

\subsection{The radial dependence of the logarithmic slope}

\begin{figure}
\plotone{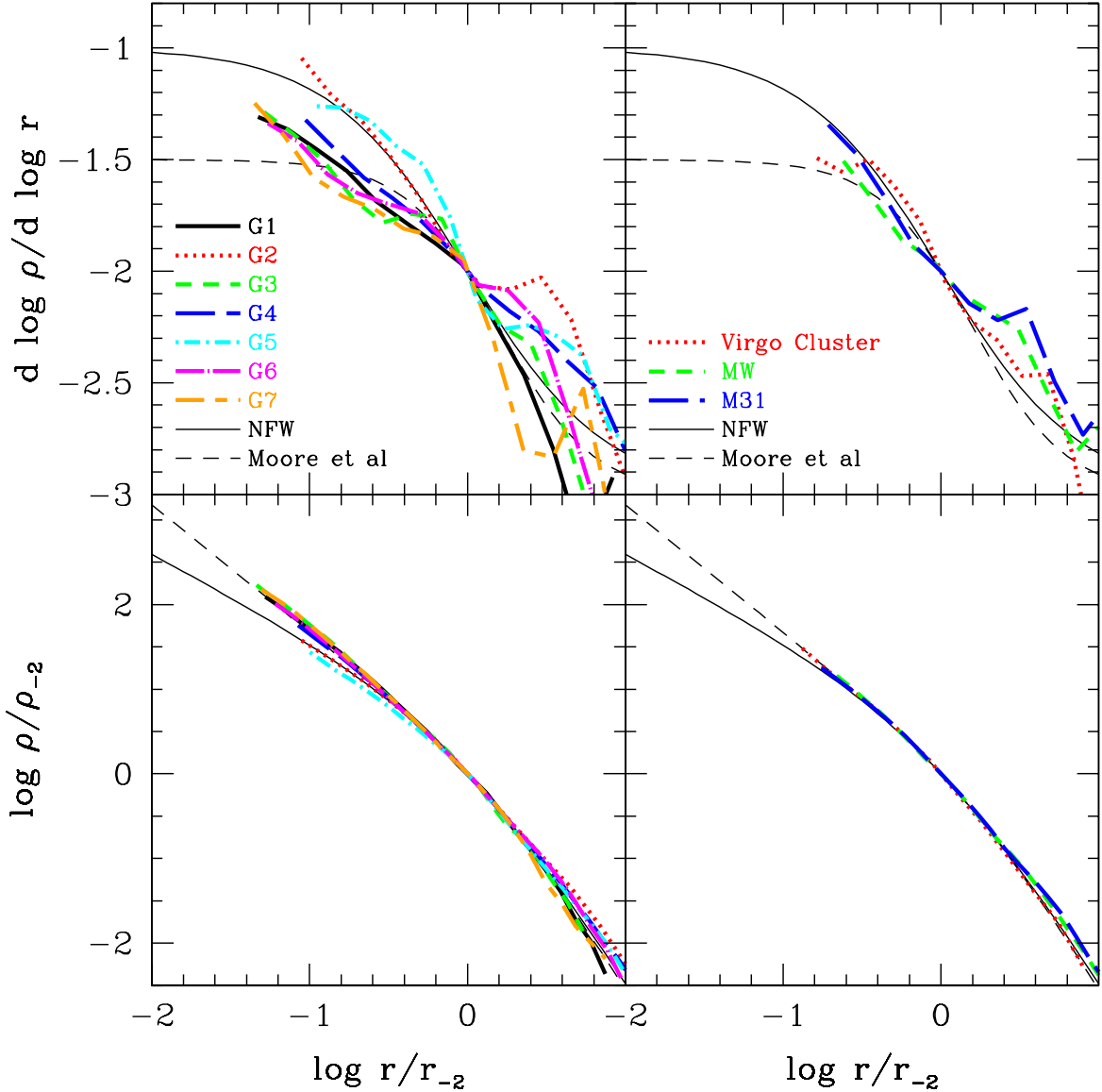}
\figcaption{{\it Upper left panel:} Logarithmic slope of the smoothed density
profile of halos simulated at our highest resolution $N_{\rm sbox}=256^3$,
plotted for $r \geq r_{\rm conv}$. Curves are scaled horizontally to the radius
$r_{-2}$, where the slope takes the isothermal value, $d\log \rho/d\log r
(r_{-2}) = -2$.  NFW and Moore \etal profiles are shown as solid and dashed
curves, respectively.  The logarithmic slope increases monotonically
with decreasing radius and there is no obvious convergence to a particular
asymptotic value of the central slope. {\it Upper right panel:} Same as upper-left
but for the SCDM Virgo cluster of \cite{GHIGNA00} and SCDM Milky Way (MW) and
M31 galaxy halos of \cite{MOORE99}.  The profiles of the two galaxy-sized halos
appear to be consistent with those of our halos.  The logarithmic slope of the
cluster halo appears to be slightly steeper than the others, fluctuating about a
value of $-1.4$ at the innermost resolved point.  {\it Lower left panel:} Halo
density profiles scaled horizontally to radius $r_{-2}$ and vertically to the
corresponding density at that radius $\rho_{-2} \equiv
\rho(r_{-2})$.  {\it Lower right Panel:} Same as lower-left but for the
\cite{GHIGNA00} and \cite{MOORE99} halos.\label{fig:dlogrho}}
\end{figure}

This view is confirmed by the radial dependence of the logarithmic slope of the
density profile $\beta(r)=- d\log \rho/d\log r$, which is shown in the top-left
panel of Figure~\ref{fig:dlogrho} for all the high-resolution runs, and compared
with the predictions of the NFW (solid line) and Moore et al (dashed line)
formulae. 

Logarithmic slopes are calculated by numerical differentiation of the
density profile, computed in radial bins of equal logarithmic width ($\Delta
\log r/r_{200} \simeq 0.2$).  The slope profiles in Figure~\ref{fig:dlogrho}
are normalized to $r_{-2}$, the radius where $\beta(r)$ takes the ``isothermal''
value of $2$.{\footnote{$r_{-2}$ is in this sense equivalent to the scale radius
$r_s$ of the NFW profile.}} In this and all subsequent figures, profiles are
shown only down to the minimum converged radius $r_{\rm conv}$.  This
corresponds typically to a radius $r_{\rm conv} \simeq 0.006~r_{200}$, or about
$1$-$2~\kpch$ for halos simulated at highest resolution (see
Table~\ref{tab:sims}).

The top left panel of Figure~\ref{fig:dlogrho} shows that halos differ
from the NFW and Moore et al formula in a number of ways:
\begin{itemize}
\item 
there is no obvious convergence to an asymptotic value of the logarithmic slope
at the centre; the profile gets shallower all the way down to the innermost
radius reliably resolved in our runs, $r_{\rm conv}$.
\item 
the slope at $r_{\rm conv}$ is significantly shallower than the asymptotic value
of $\beta_0=1.5$ advocated by \cite{MOORE99}.  The shallowest value measured
at $r_{\rm conv}$ is $\beta \simeq 1$, and the average over all \nhalos halos is
$\beta \simeq 1.2$.
\item 
Most halo profiles become shallower with radius more gradually than predicted by
the NFW formula; at $r\sim 0.1~r_{-2}$ the average slope is $\sim -1.4$, whereas
NFW would predict $\sim -1.18$.  The NFW density profile turns over too sharply
from $\rho \propto r^{-3}$ to $\rho \propto r^{-1}$ compared to the simulations.
\end{itemize}

In other words, the Moore et al profile appears to fit better the inner regions
of the density profile of some $\Lambda$CDM halos (see bottom-left panel of
Figure~\ref{fig:dlogrho}) {\it not} because the inner density cusp diverges as
steeply as $\beta_0=1.5$, but rather because its logarithmic slope becomes
shallower inwards less rapidly than NFW. 

It is important to note as well that there is significant scatter from halo to
halo, and that two of the seven density profiles are actually fit better by the
NFW formula. Are these global deviations from a ``universal'' profile due to
substructure? We have addressed this question by removing substructure from all
halos and then recomputing the slopes. Substructure is removed by first
computing the local density at the position of each particle, $\rho_i$, using a
spline kernel similar to that used in Smoothed Particle Hydrodynamics (SPH)
calculations\footnote{See {\tt
http://www-hpcc.astro.washington.edu/tools/smooth.html}}. Then, we remove all
particles whose densities are more than 2 standard deviations above the
spherically-averaged mean density at its location. (The mean and standard
deviation are computed in bins of equal logarithmic width, $\Delta \log
r/r_{200} \simeq 0.01$). The procedure is iterated until no further particles
are removed. The remaining particles form a smoothly distributed system that
appears devoid of substructure on all scales.  We find that density profiles are
smoother after the removal of substructure but that most of the variation in the
overall shapes of the profiles remains.  We conclude that the presence of
substructure is not directly responsible for the observed scatter in the shape
of halo density profiles.

\subsection{Comparison with Other Work}
\label{sec:compben}

Are these conclusions consistent with previous work? To explore this issue, we
have computed the logarithmic slope profile of three CDM halos run by Moore and
collaborators.  The halos we have re-analyzed are the Milky Way- and M31-like
galaxy halos of the Local Group system from \cite{MOORE99} and the LORES version
of the ``Virgo'' cluster halo from \cite{GHIGNA00}.  The $z \simeq 0.1$ output
of the Local Group simulation was provided to us by the authors, whereas the
Virgo cluster was re-run using initial conditions available from Moore's
website{\footnote{{\tt http://www.nbody.net}. We note that all of these runs
were evolved in an $\Omega_0=1$ cosmogony, rather than the $\Lambda$CDM scenario
we adopt in this paper.}}. The Virgo cluster run used the same N-body code as
the original simulation ({\tt PKDGRAV}) but was run with a fixed number of
timesteps (12800). A run with 6400 timesteps was also carried out and no
differences in the mass profiles were detected. The number of particles within
the virial radius is $1.2 \times 10^6$, $1.7 \times 10^6$, and $5.0 \times
10^5$, for the Milky Way (MW), M31 and LORES Virgo cluster halos, respectively.

Figure~\ref{fig:dlogrho} shows the logarithmic slope (upper right panel) and
density (lower right panel) profiles corresponding to these halos, plotted down
to the minimum converged radius $r_{\rm conv}$. No major differences between
these simulations and ours are obvious from these panels.  It is clear, for
example, that at the innermost converged point, the slope of the density profile
of the two Local Group halos is significantly shallower than $r^{-1.5}$, and
shows no signs of having converged to a well defined power-law behaviour. There
is some evidence for ``convergence'' to a steep cusp ($r^{-1.4}$) in the LORES
Virgo cluster simulation but the dynamic range over which this behaviour is
observed is rather limited. The Virgo cluster run thus appears slightly unusual
when compared with other systems in our ensemble. Although our reanalysis
confirms the conclusion of \cite{MOORE98,MOORE99} that this system appears to
have a steeply divergent core, this does not seem to be a general feature of
$\Lambda$CDM halos.

Our results thus lend support to the conclusions of \cite{KLY01}, who
argues that there is substantial scatter in the inner profiles of cold dark
matter halos. Some are best described by the NFW profile whereas others are
better fit by the Moore et al formula, implying that studies based on a single
halo might reach significantly biased conclusions. 

Finally, we note that deviations from either fitting formula in the radial range
resolved by the simulations, although significant, are small. Best NFW/Moore et
al fits are typically accurate to better than $\sim 20\%$ in circular velocity
and $\sim 40\%$ in density, respectively. We discuss in a companion paper
\citep{NAVPREP03} the constraints placed by our simulations on extrapolations of
these formulae to the inner regions as well as on the true asymptotic inner
slope of $\Lambda$CDM halo density profiles.

\section{Halo Circular Velocity Profiles and LSB Rotation Curves}
\label{sec:rotcurve}

As discussed in \S~\ref{sec:intro}, an important discrepancy between the
structure of CDM halos and the mass distribution in disk galaxies inferred from
rotation curves has been noted repeatedly in the literature over the past decade
\citep{MOORE94,FLORES94,BURKERT95,MCGAUGH98,MOORE99,VDB00,COTE00,BLAIS01,VDB01,JIMENEZ03}.
In particular, the shape of the rotation curves of low surface brightness (LSB)
galaxies has been identified as especially difficult to reconcile with the
``cuspy'' density profiles of CDM halos.

Given the small contribution of the baryonic component to the mass budget in
these galaxies, the rotation curves of LSB disks are expected to trace rather
cleanly the dark matter potential, making them ideal probes of the inner
structure of dark matter halos in LSBs. Many of these galaxies are better fit by
circular velocity curves arising from density profiles with a well defined
constant density ``core'' rather than the cuspy ones inferred from simulations,
a result that has prompted calls for a radical revision of the CDM paradigm on
small scales \citep[see e.g.,][]{SS00}
 
It is important, however, to note a number of caveats that apply to the LSB
rotation curve problem.
\begin{itemize}
\item
Many of the early rotation curves where the disagreement was noted were unduly
affected by beam smearing in the HI data \citep{SWATERS00}. For example,
\cite{VDB00} argue that, once beam smearing is taken into account, essentially
all HI LSB rotation curves are consistent with cuspy halo profiles. The
observational situation has now improved substantially thanks to
higher-resolution rotation curves obtained from long-slit $H_{\alpha}$
observations \citep[see, e.g.,][]{MCGAUGH01,DEBLOK01A,SWATERS00,SWATERS03}. We
shall restrict our analysis to these rotation curves in what follows.
\item
Strictly speaking, the observational disagreement is with the fitting formulae,
rather than with the actual structure of simulated CDM halos.  As noted in the
previous section, there are systematic differences between them, so it is
important to confirm that the disagreement persists when LSB rotation curves are
contrasted directly with simulations.
\item
Finally, it must be emphasized that the rotation curve problem arises when
comparing rotation speeds of gaseous disks to the spherically-averaged circular
velocity profiles of dark matter halos. Given that CDM halos are expected to be
significantly non-spherical \citep{BARNES87, WARREN92,JING95,THOMAS98,JING02},
some differences between the two are to be expected. It is therefore important
to use the full 3D structure of CDM halos to make predictions regarding the
rotation curves of gaseous disks that may be compared directly to
observation. We shall neglect this complex issue in this paper, but plan to
explore in detail the rotation curves of gaseous disks embedded in such
asymmetric potentials in future papers of this series.
\end{itemize}

We may avoid many of these uncertainties by comparing directly the circular
velocity profiles of our simulated halos with the observational data.  This
procedure has the advantage of retaining the diversity in the shapes of halo
profiles that is often lost when adopting a simple analytic fitting formula.  In
addition, we consider circular velocity profiles only down to the innermost
converged radius, thereby eliminating uncertainties about the reliability of the
profile at very small radii.

We begin the analysis by emphasizing the importance of taking into account the
changes in the central halo mass profile induced by accretion events. Indeed,
these may trigger and sustain departures from the ``average'' profile that may
be detectable in the rotation curves of embedded gaseous disks. We shall then
describe a simple characterization of rotation curve shapes that may be applied
to both observational and simulation data. This enables a direct and
quantitative assessment of the ``cusp'' versus ``core'' problem as it applies to
the most recent LSB datasets.

\subsection{Evolution of the Inner Mass Profile}
\label{sec:vcvar}

Systematic---and at times substantial---changes in the inner circular velocity
profile are induced by accretion events during the assembly of the halo, even
when such these events might contribute only a small fraction of mass to the
inner regions. These transients may increase substantially the scatter in the
shape of the $V_c$ profiles and they ought to be taken into account when
comparing with observation.

\begin{figure}
\centerline{\epsfig{figure=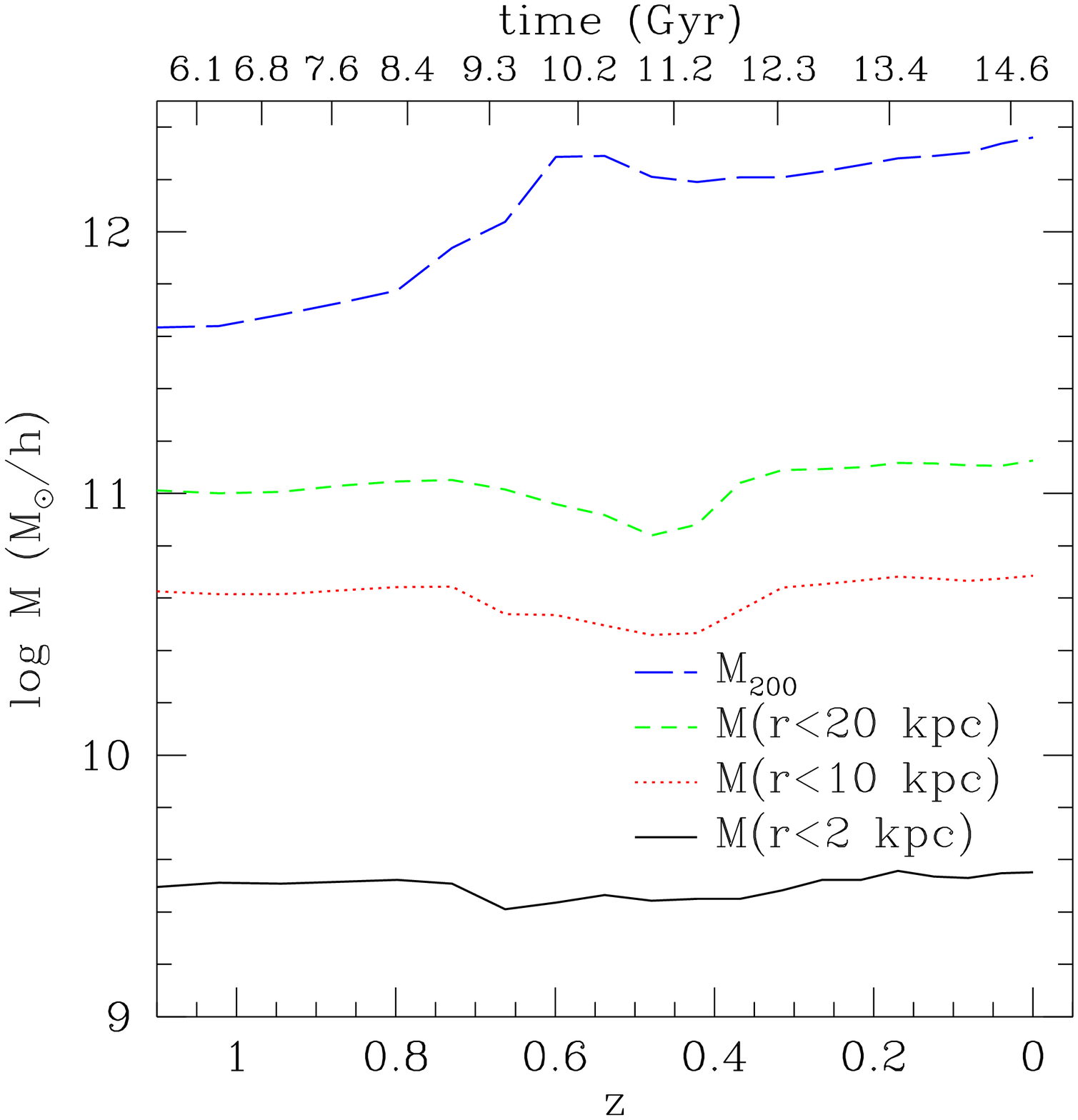,width=0.6\linewidth}}
\centerline{\epsfig{figure=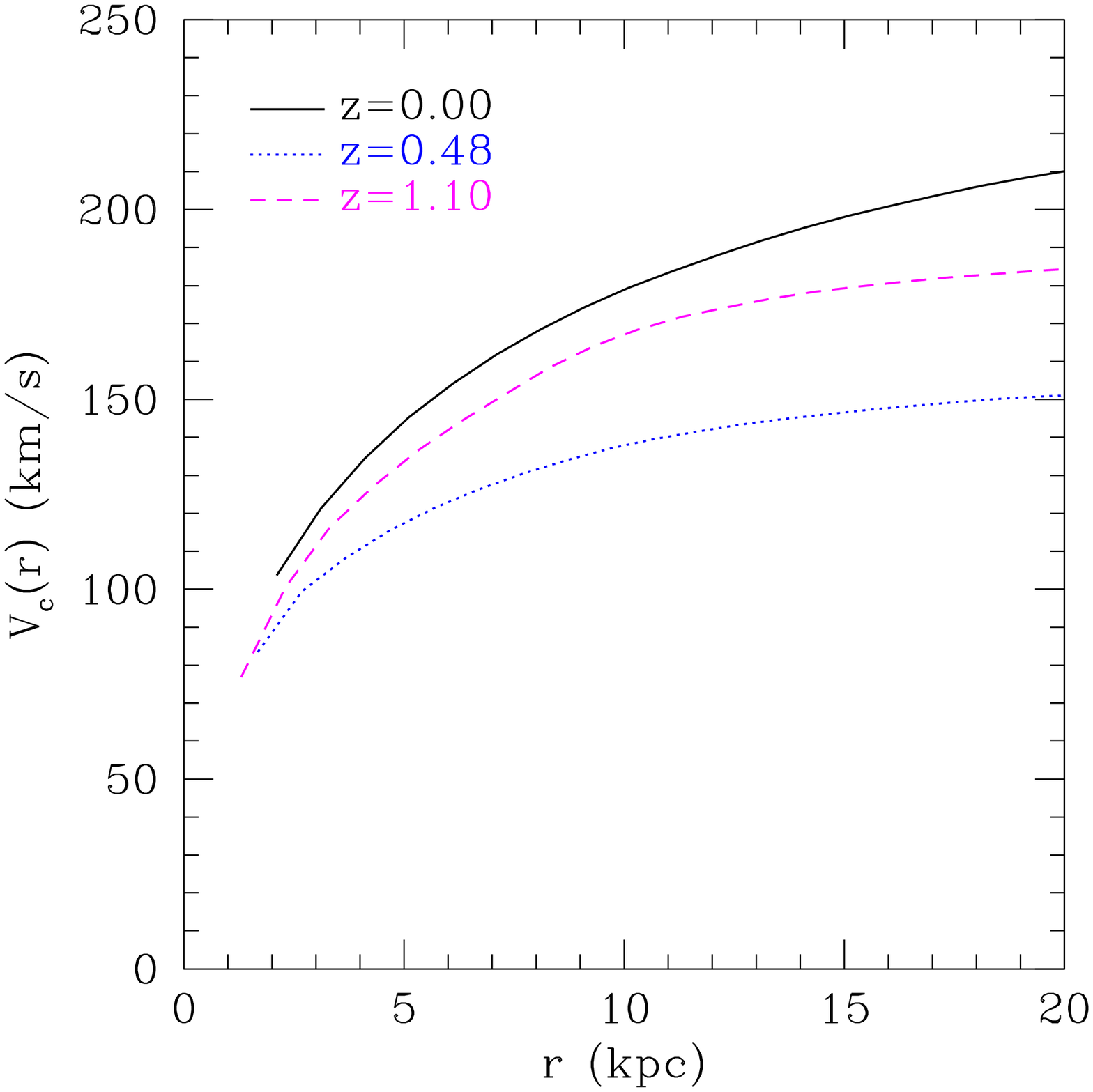,width=0.6\linewidth}}
\figcaption{{\it Top panel:} Mass within $r=2$, $10$, and $20$ kpc (physical)
and $r_{200}$ for halo G1/$256^3$ as a function of redshift (age of Universe in
Gyr) on bottom (top) axis.  The mass within $20~\kpc$ undergoes significant
fluctuations in response to a merger at $z \simeq 0.7$.  {\it Bottom panel:} The
inner circular velocity profile before ($z=1.10$), during ($z=0.48$), and after
($z=0$) the merger.  The shape of the $V_c$ profile is noticeably altered by the
effects of the infalling substructure.\label{fig:cmshell}}
\end{figure}

This is illustrated in Figure~\ref{fig:cmshell}, which shows the evolution of
the mass and circular velocity profiles of halo G1. The top panel of this figure
shows the evolution of the mass enclosed within $8$, $10$, $20$ kpc (physical),
and $r_{200}$, as a function of redshift. Although the mass inside $20~\kpc$
increases by less than $\sim 25\%$ since $z=1$, there are significant ($\sim
50-60\%$) fluctuations during this time caused by the tidal effects of orbiting
substructure and accretion events.  Most noticeable is a major merger at $z
\simeq 0.7$, which affects the mass profile down to the innermost reliably
resolved radius, $\simeq 2~\kpc$.

The effect of these fluctuations on the circular velocity profile is shown in
the bottom panel of Figure~\ref{fig:cmshell}.  Here we show the inner $20 \kpc$
of the circular velocity profile before ($z=1.1$), during ($z=0.48$) and after
($z=0$) a major accretion event.  Substantial changes in the shape of the $V_c$
profile are evident as the halo responds to the infalling substructure.  Note
that the changes persist over timescales of order $\gsim 1$ Gyr, exceeding the
circular orbital period at $r=2$, $10$, and $20$ kpc ($\sim 0.13$, $0.34$, and
$0.58$ Gyr, respectively).  These relatively long lasting changes thus would
likely be reflected in the dynamics of a disk present at the centre of the halo.

\subsection{LSB Rotation Curves}
\label{sec:compobs}

Could the evolutionary effects discussed above be responsible, at least in part,
for the constant density cores inferred from the rotation curves of LSB and
dwarf galaxies?  Since it is nearly impossible to tailor a simulation to
reproduce individual galaxies in detail, it is important to adopt a simple
characterization of the rotation curves that allows for a statistical assessment
of the disagreement between halo $V_c$ profiles and observation. We have thus
adopted a three-parameter fitting formula commonly used in observational work to
describe optical rotation curves
\citep{COURTEAU97}:
\begin{equation}
V(r) = \frac{V_0}{(1+x^\gamma)^{1/\gamma}}.
\label{eq:multifit}
\end{equation}

Here $V_0$ is a velocity scale, $x=r_t/r$, where $r_t$ is a scale radius, and
the dimensionless parameter $\gamma$ describes the overall shape of the curve.
The larger the value of $\gamma$ the sharper the turnover from the ``rising'' to
the ``flat'' region of the velocity curve.  Eq.~\ref{eq:multifit} is flexible
enough to accommodate the shape of essentially all rotation curves in the
samples we consider here.  We note that this formula has three\footnote{An
additional factor of $(1+x)^\beta$ was used by \cite{COURTEAU97} to improve fits
to $\sim 10\%$ of the galaxies in his sample that exhibit a drop-off in the
outer part of the rotation curve. For simplicity, we have not included this
parameter in our fits.} free parameters, one more than the NFW profile.

We have applied this fitting formula to the HI/H$\alpha$ rotation curve datasets
of \cite{DEBLOK01}\footnote{\tt http://www.astro.umd.edu/$\sim$ssm/data},
\cite{DEBLOK02} \footnote{\tt ftp://cdsarc.u-strasbg.fr/cats/J/A+A/385/816}, and
\cite{SWATERS03} \footnote{\tt http://www.robswork.net/data}, hereafter B01, B02, and S03,
respectively. Fits to the rotation curves and $V_c$ profiles are obtained
through straightforward $\chi^2$-minimization, adopting the error estimates
provided by the authors.

The B01 sample consists of 26 LSB galaxies, the B02 sample consists of 24 LSB
galaxies, and the S03 sample contains 10 dwarf galaxies and 5 LSB galaxies. 
The smoothed rotation curves of B01 were derived by folding approaching and
receding velocities about the centre of the galaxy (defined by the peak in the
continuum emission) and using a spline-fitting procedure followed by rebinning
to a bin width of 2''.  Error estimates were calculated as the quadratic sum of
an observational error component caused by measurement uncertainties in the raw
data points, and an additional error component due to differences between the
approaching and receding velocities in the bin, as well as noncircular motions
(defined as the difference between the mean velocity and the velocity of the
spline fit). The rotation curve data provided by B02 are unsmoothed, with errors
which include measurement, inclination, and asymmetry uncertainties. The
rotation curves of S03 were derived by averaging receding and approaching
velocity data in 2'' bins; error estimates were defined by half the quadratic
sum of the average error of points in the bin and half the difference between
the maximum and minimum velocities in the bin.

The fit parameters for the combined set of 65 rotation curves are given in
Table~\ref{tab:multifit}. In this table, $r_{\rm max}$ refers to the radius
where the rotation curve reaches the maximum, $V_{\rm max}$. The outermost
radius with reliable data is listed as $r_{\rm
outer}$. Figure~\ref{fig:multifitsamp} shows a selected sample of rotation
curves, together with the best fits obtained with eq.~\ref{eq:multifit}.  The
top, middle, and bottom rows include galaxies from the S03, B02, and B01
samples, respectively, arranged from left to right in order of increasing value
of the shape parameter $\gamma$.

\begin{figure}
\plotone{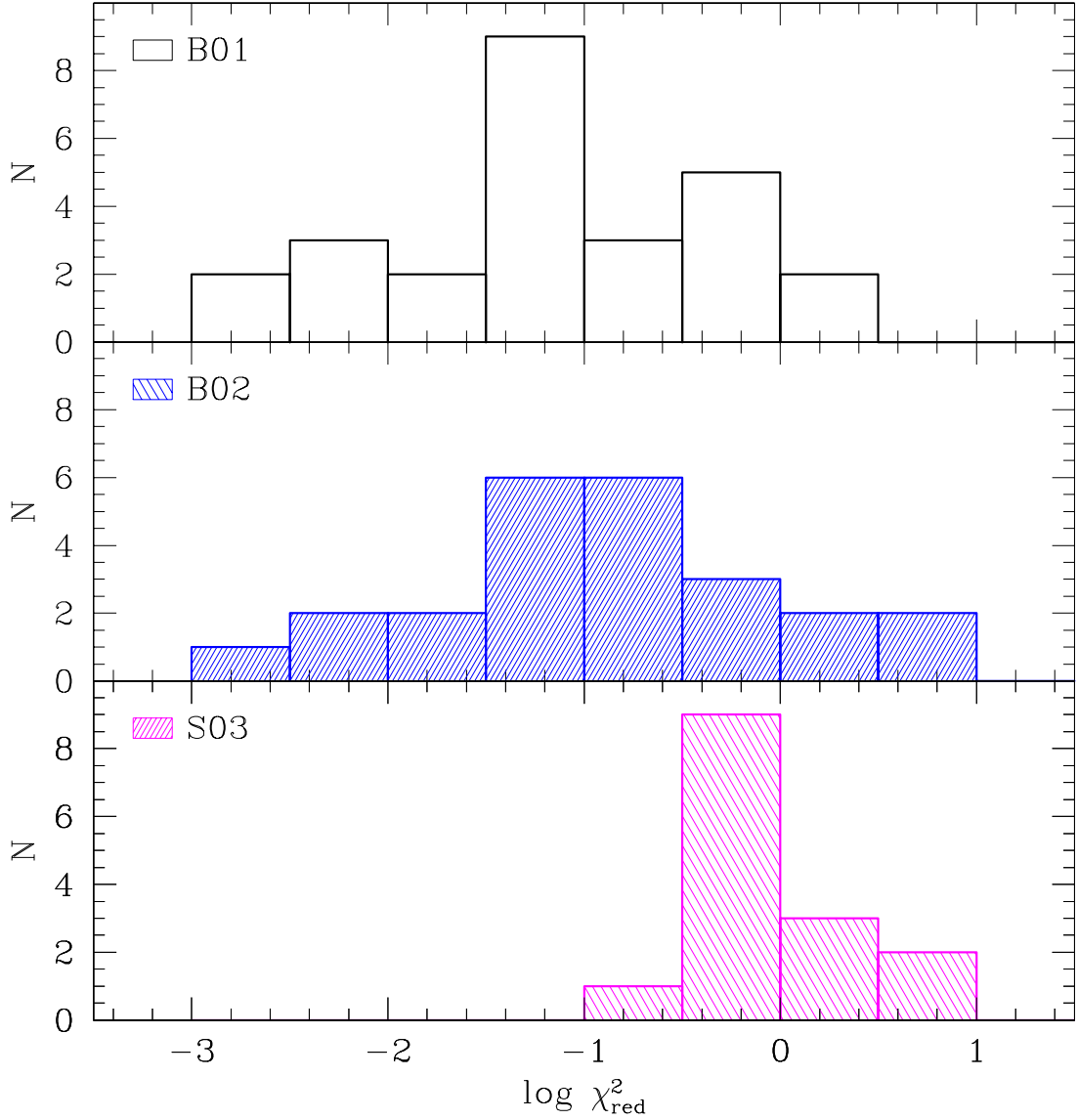}
\figcaption{
Distribution of reduced $\chi^2$ values corresponding to best fits using the
multi-parameter fitting formula given by eq.~\ref{eq:multifit} for the three
rotation curve datasets (B01, B02, S03).  Note that the B01 and B02 $\chi^2_{\rm
red}$ distributions peak at unrealistically low values, signalling the presence
of significant correlation between neighbouring points in the rotation curves of
these samples. The S03  $\chi^2$ distribution peaks at a higher value,
emphasizing the various assumptions of different authors when deriving rotation
curves from raw data (see also Figure~\ref{fig:multifitcmp}).
\label{fig:chi2hist}}
\end{figure}

It is important to note that the rotation curves of B02 and S03 differ
significantly from those of B01. The smoothing applied to the B01 dataset is
clear in this figure, especially when compared with the less-processed B02 and
S03 datasets. Because the B01 curves have been smoothed, the individual values
quoted for the rotation speed are correlated, and the error estimates lack clear
statistical meaning. This is confirmed by the extremely low values of the formal
reduced $\chi^2$ obtained for the best fits with eq.~\ref{eq:multifit} (see
Table~\ref{tab:multifit} and Figure~\ref{fig:chi2hist}): $16$ out of $26$ B01
galaxies have $\chi^2_{\rm red} <0.1$ (and $5$ have $\chi^2_{\rm red} <0.01$),
whereas all galaxies in S03's sample have $\chi^2_{\rm red} >0.1$. This clearly
advises against using $\chi^2$ as a goodness-of-fit measure intended to rule out
a particular model.

\subsection{LSB rotation curve shapes}

\begin{figure}
\plotone{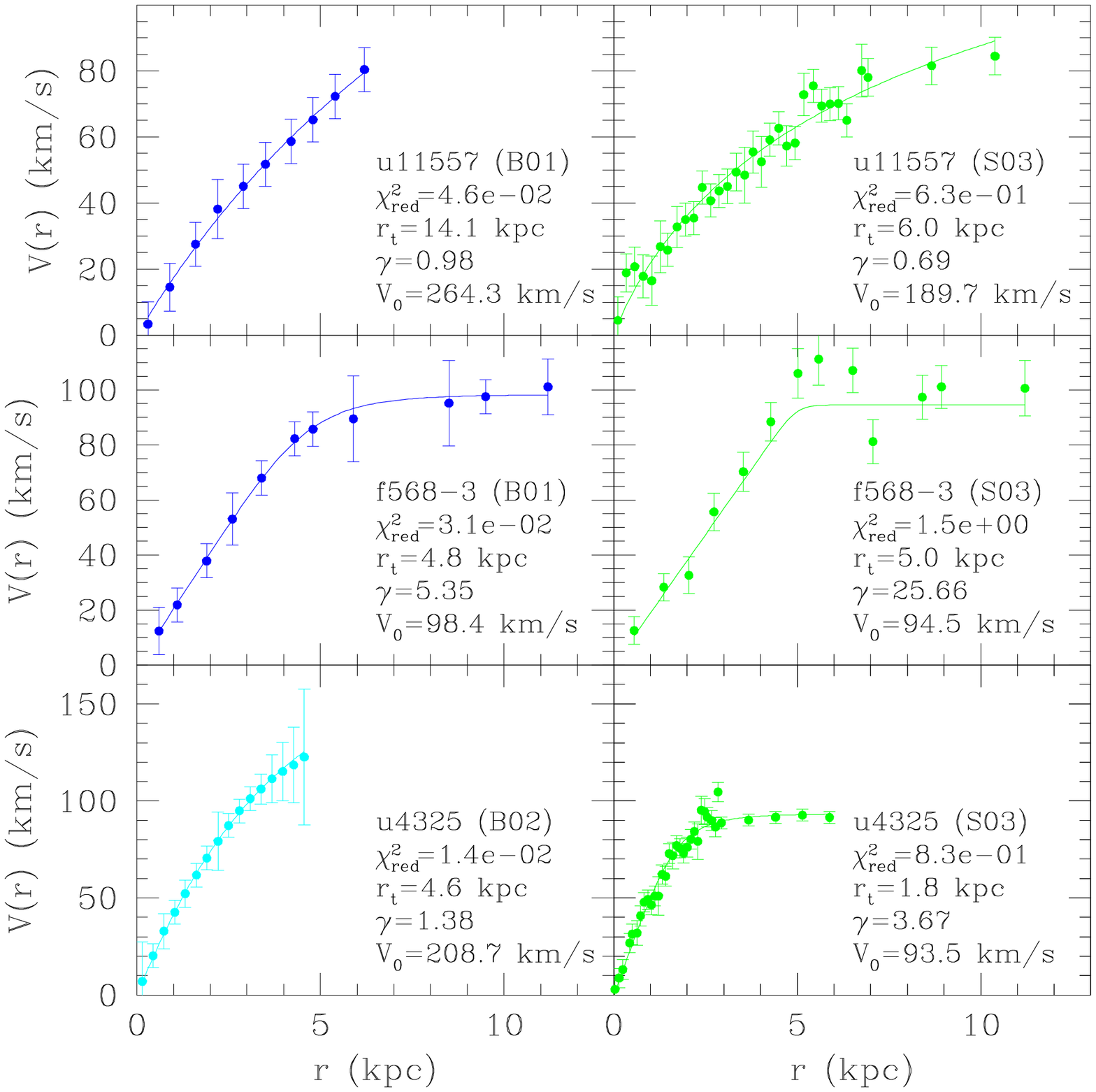}
\figcaption{
Rotation curves of UGC 11557, F568-3, and UGC 4325 derived by either
\cite{DEBLOK01, DEBLOK02} (B01 and B02, left panels) or \cite{SWATERS03} (S03,
right panels). Solid curves show fits to the rotation curves using
eq.~\ref{eq:multifit}.  Values of the three fitting parameters $\gamma$, $r_t$,
and $V_0$ are listed in each panel along with the corresponding reduced
chi-squared, $\chi_{\rm red}^2$. This figure illustrates the effect on the
derived rotation curves due to different assumptions and methodology adopted by
various authors.  In the case of UGC 4325, for example, the rotation curve
derived by B02 continues to rise out to the last measured point at $V\simeq
120~\kms$, whereas the the S03 curve flattens off at $V\simeq
90~\kms$.\label{fig:multifitcmp}}
\end{figure}

The analysis procedures used by various authors induce significant differences
in the rotation curves derived from the data, as illustrated by comparing the
rotation curves of UGC 11557, and F568-3, the two galaxies common to the B01 and
S03 samples, and UGC 4325, the only galaxy common to the B02 and S03 datasets.
(Figure~\ref{fig:multifitcmp}).  The rotation curve of UGC 11557
(Figure~\ref{fig:multifitcmp}, upper left panel) presented by B01 extends out to
only $r \simeq 6~\kpc$ and rises with a nearly constant slope out to this radius.  The
best fit multi-parameter function has a scale radius $r_t = 14.1~\kpc$, much
greater than the outermost radius of the observations, $r_{\rm outer} =
6.2~\kpc$, and a low value of $\gamma = 0.98$. The S03 rotation curve
(Figure~\ref{fig:multifitcmp}, upper right panel) for the same galaxy extends
out to $r \simeq 10~\kpc$ and appears to level off in the last two data points.
The scale radius of the fit to this curve is $r_t = 6~\kpc$: well within the
outermost radius of the observations $r_{\rm outer} = 10.4~\kpc$, and different
by a factor of two from that of the B01 curve.  Although the value of the shape
parameter $\gamma$ is actually lower for the S03 ($\gamma=0.69$), it is clear
that additional data points in the flat part of the curve, or smaller error bars
on the last few points, would force the fit to higher values of $\gamma$.  This
would also result in fits with a much lower value of the asymptotic velocity
$V_0$, which presently has a value much greater than the maximum velocity
measured in either of the B01 or S03 rotation curves.

In the case of F568-3, B01's rotation curve (Figure~\ref{fig:multifitcmp},
middle left panel) rises monotonically out to the last point and as a result,
the multi-parameter function fits quite well ($\chi_{\rm red}^2 = 0.031$).  The
S03 rotation curve extends out to the same outermost radius as the B01 curve and
both curves are consistent with one another within the observational errors.
Due to the different methods of estimating velocities, however, the B01 rotation
curve is much smoother than that of S03.  The S03 rotation curve
(Figure~\ref{fig:multifitcmp}, middle right panel) rises to a maximum at $r
\simeq 6~\kpc$ then falls to $V_c \simeq 0.9~V_{\rm max}$ at the outermost radius
$r_{\rm outer} = 11.2~\kpc$.  The fit to this curve is further complicated by a
deviant data point at $r= 7~\kpc$ with a velocity, $V_c \simeq 0.7~V_{\rm max}$, much
lower than that of neighbouring points.  It is obvious that no simple fitting
function can provide a good fit to this rotation curve.  The best fit
multi-parameter function has a poor goodness-of-fit statistic $\chi_{\rm red}^2
= 1.5$ and is characterized by an inordinately sharp turnover ($\gamma =
25.7$). We also note that the error bars in the B01 and S03 data are comparable
in size ($\simeq \pm 8~\kms$) except near the radius of the discrepant point in
the S01 curve, where error bars in the B01 data are twice as large ($\simeq \pm
16~ \kms$).  Clearly, the shape of the rotation curve as parameterized by
fitting functions like the one given by eq.~\ref{eq:multifit} is strongly
influenced by the sensitivity of the observations, as well as by the method used
to determine the rotation curve from the raw data.

Perhaps the most interesting case is the rotation curve of UGC 4325.  The B02
and S03 versions of the rotation curve appear qualitatively different and are
not consistent with one another within the error bars as presented.  The B02
rotation curve rises monotonically out the last measured point, whereas the S03
curve flattens off sharply to a well-defined asymptotic value.  Fits to the B02
and S03 rotation curves yield significantly different values for all three
fitting parameters.  The sharp turnover of the S03 curve results in a relatively
large value of $\gamma = 3.67$ compared to $\gamma=1.38$ for the fit to the B02
rotation curve.  In addition, the asymptotic velocity $V_0$ of the B02 fit is
more than twice that of the S03 fit.  Of even greater concern is the maximum
velocity one might infer for UGC 4325 from each version of the rotation curve.
In the case of the B02 curve, the maximum velocity is undetermined but certainly
appears to be greater than $V(r_{\rm outer}) \simeq 122~\kms$ since the curve is
still rising at the outermost data point.  According to the S03 curve, however,
the maximum velocity is robustly determined, equal to the asymptotic velocity of
the fit, $V_0 \simeq 94~\kms$.  The maximum velocity one infers for UGC 4352
therefore differs by at least 30\% depending on which rotation curve is used.
This may have important consequences for the calculation of scaling parameters such
as the central densities of galaxies, a point we return to in \S\ref{ssec:conc}.

\begin{figure}
\plotone{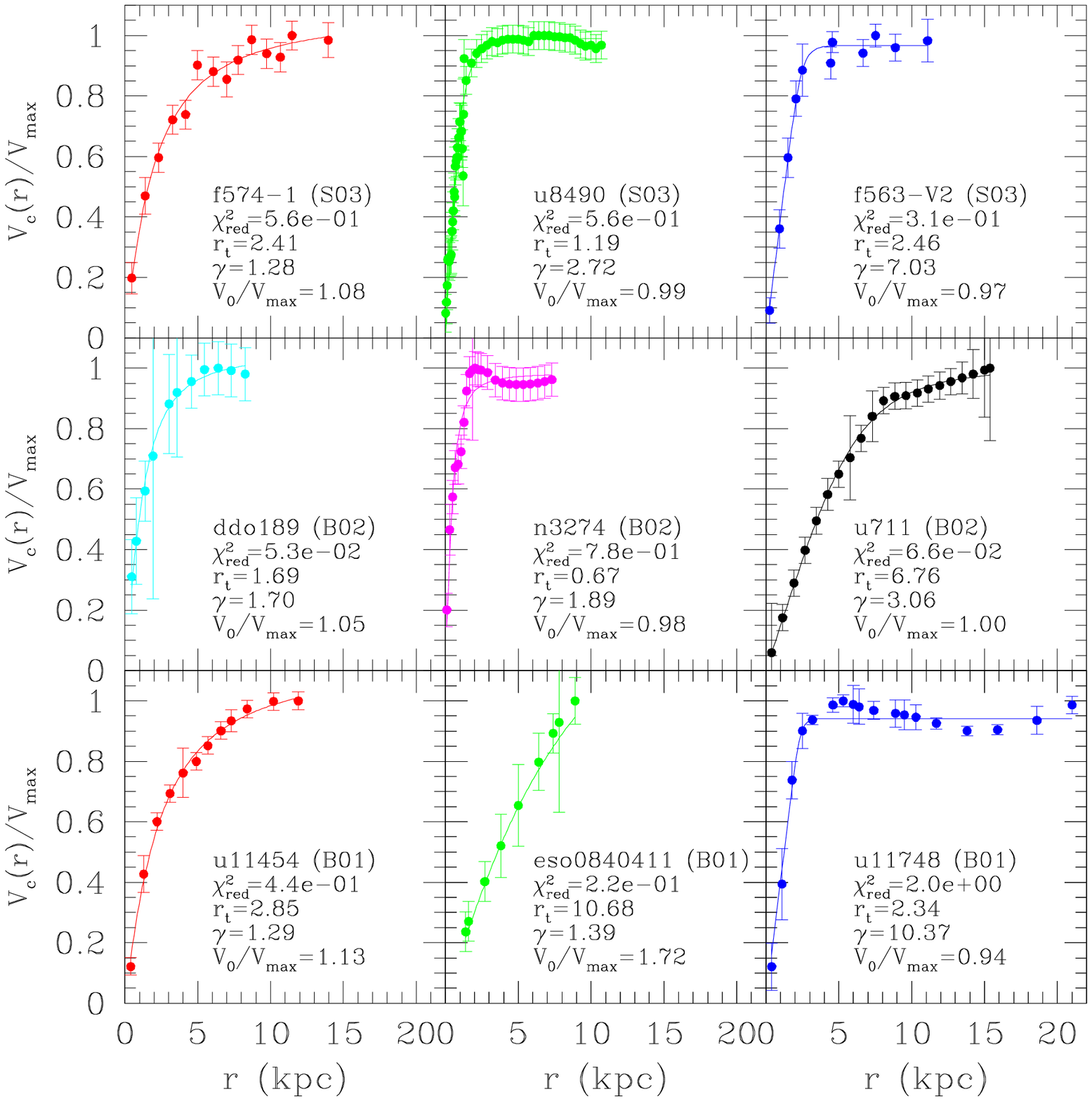}
\figcaption{LSB rotation curves from the datasets of
\cite{DEBLOK01} (B01), \cite{DEBLOK02} (B02) and \cite{SWATERS03} (S03).  Solid
curves show best fits using the \cite{COURTEAU97} fitting formula given by
eq.~\ref{eq:multifit}. Best fits with $\gamma \sim 1$ correspond to rotation
curves that rise and turn over gently as a function of radius.  Curves with
$\gamma > 2$ feature a much sharper transition from the rising to the flat part
of the curve. See text for full discussion.
\label{fig:multifitsamp}}

\end{figure}

The spread in the best-fit parameters listed in the panels of
Figure~\ref{fig:multifitcmp} is again a sobering reminder that a fair amount of
non-trivial processing is involved when deriving rotation curves from raw
data. Clearly, the disagreement between authors is somewhat worrying, and it
limits the general applicability of conclusions inferred from individual galaxy
studies. In spite of this, there appears to be broad statistical consistency
between the rotation curve parameters derived by B01, B02 and S03.  This is
illustrated in the top panel of Figure~\ref{fig:gammahist}, where we show the
distribution of best-fit $\gamma$ values obtained for each sample.  Each
histogram in this figure is normalized to the total number of systems in each
sample for ease of comparison.  All three rotation curve datasets are broadly
consistent with each other; in each case most ($70\% \pm 5\%$) LSB rotation
curves are characterized by a value of $\gamma < 2$. These are typically
gently-rising curves which turn over gradually as they approach the maximum
asymptotic rotation speed, as shown, for example, by F574-1, and UGC 11454 in
Figure~\ref{fig:multifitsamp}. A significant ($\simeq 30\%$) number of `outlier'
galaxies with $\gamma \gsim 2.5$, however, populate the tail of the combined
B01+B02+S03 distribution (see bottom panel of Figure~\ref{fig:gammahist}). These
are galaxies whose rotation curves feature a much ``sharper'' transition from
the rising to flat part, as shown, for example, by F563-V2 and UGC 11748 in
Figure~\ref{fig:multifitsamp}.

\subsection{Halo circular velocity profile shapes}

\begin{figure}
\plotone{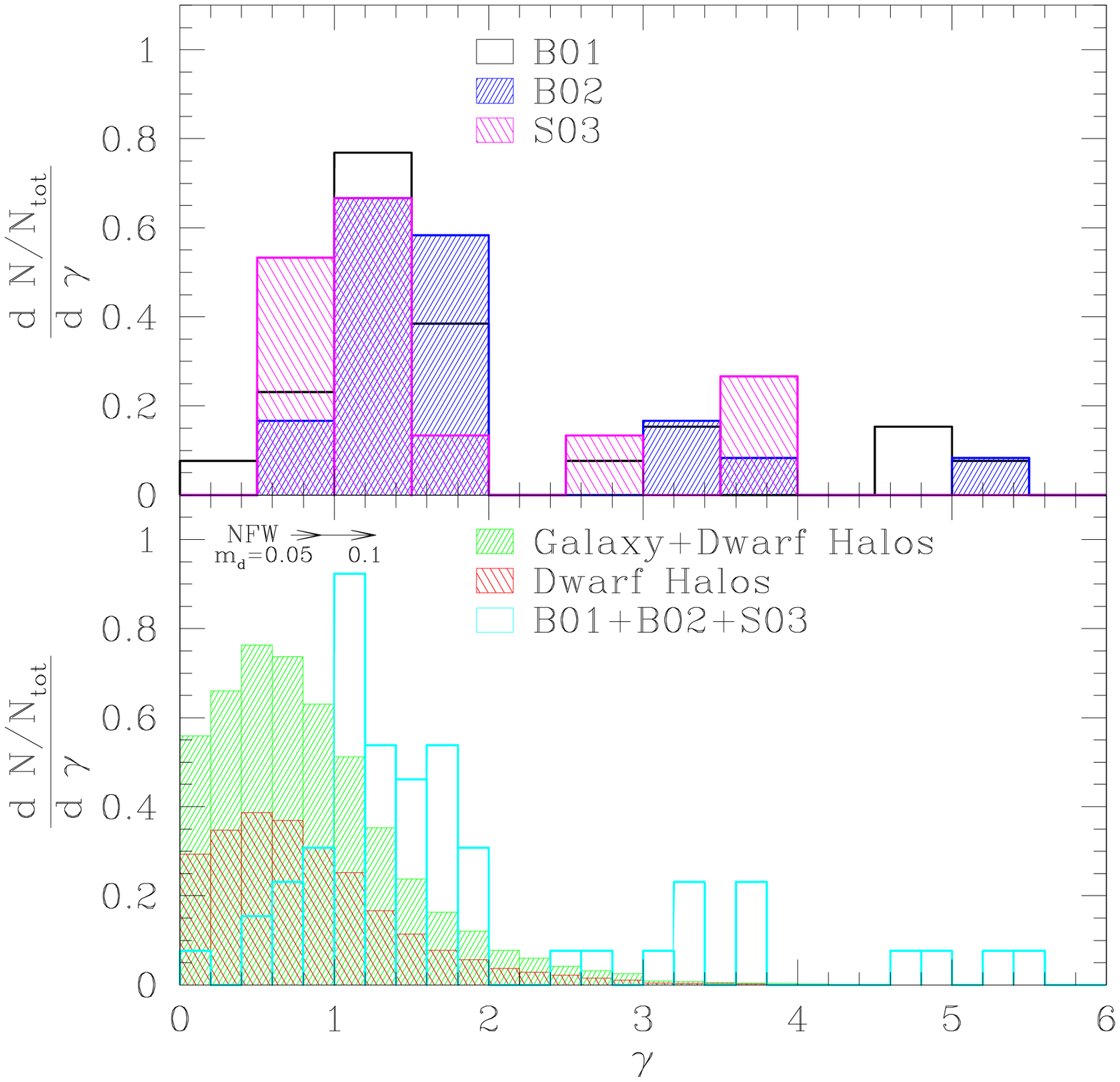}
\figcaption{ 
{\it Top panel:} Distribution of best-fit $\gamma$ values obtained for galaxies
in the samples of \cite{DEBLOK01} (B01), \cite{DEBLOK02} (B02), and
\cite{SWATERS03} (S03). Note that most galaxies in all three samples cluster
around $\gamma=1.2$, but that there are a significant number of outliers with
$\gamma \gsim 2.5$. These define a population of galaxies that seems to be
distinct from the bulk of the sample.  {\it Bottom panel:} Combined
observational sample compared with the halo $\gamma$ distribution after
convolution with an error distribution similar to that corresponding to observed
rotation curve fits.  Arrows show the change in $\gamma$ caused by the addition
of an exponential disk to an NFW halo with concentration $c=10$.  The mass of
the disk, $m_d$, is given in units of the halo mass.  Its exponential scale
length is computed assuming that the spin parameter of the halo is
$\lambda=0.1$, and that the specific angular momentum of the disk is the same as
that of the halo \citep{MO98}.  The magnitude of the correction (shown by
horizontal arrows) suggests that the $\gamma$ distribution of halos might
actually be consistent with that of the bulk of galaxies, but apparently fails
to account for the $\gamma\gsim 2.5$ tail in the rotation curve
distribution. See text for further details.
\label{fig:gammahist}}
\end{figure}

The bottom panel of Figure~\ref{fig:gammahist} shows the $\gamma$ distribution
obtained by fitting eq.~\ref{eq:multifit} to the $V_c$ profile of all dwarf and
galaxy-sized halos. In order to consider the various dynamical instances of a
halo, we have included in the analysis about $20$ different outputs for each
system, spanning the redshift range $1 \lsim z \leq 0$, giving a total of 266
halo profiles.  We calculated the $V_c$ profile at each redshift in bins
$1~\kpc$ (physical) in width for the galaxy halos and $0.2~\kpc$ (physical) in
width for the dwarf halos, starting at the innermost reliably resolved radius
$r_{\rm conv}$.  The $V_c$ profiles were fit out to $r=20~\kpc$ (physical) for
the galaxy halos and $r=8~\kpc$ (physical) for the dwarf halos, although we note
that the fitting parameters are fairly insensitive to the region fitted provided
that the curvature of the profile is well resolved.  Since no formal error bars
exist for the halo profiles (Poisson errors are negligible for the numbers of
particles in these halos), we assign a uniform error of $\pm 1~\kms$ to all
points.

The bottom panel of Figure~\ref{fig:gammahist} shows the distribution of
$\gamma$ values obtained for the halos, after convolution with the typical
uncertainty in $\gamma$ derived from fits to the observed rotation curves (an
asymmetric Gaussian with $\sigma_{\gamma +} = 1.0$, $\sigma_{\gamma -} = 0.5$).
The $\gamma$ distribution of all galaxy and dwarf halos, that of the dwarf halos
only, and that of the combined B01, B02 and S03 sample are shown as the green,
red and open histograms, respectively.  The green and red histograms are
normalized to the total number of halos, and the open histogram is normalized to
the total number of galaxies in the combined observational dataset.  As shown in
Figure~\ref{fig:gammahist}, the convolved halo $\gamma$ distribution peaks at
$\gamma \simeq 0.6$, and has a dispersion of order $\sim 0.6$. For illustration,
the three G1 $V_c$ profiles shown in the bottom panel of
Figure~\ref{fig:cmshell} have $\gamma = 0.73$, $0.65$, and $0.48$ at $z=1.1$,
$0.48$, and $0$, respectively.  There is no significant difference between the
galaxy and dwarf halo $\gamma$ distributions.

\subsection{Comparison}

How does the distribution of rotation curve shapes compare with that of halo
circular velocity profiles?  The bottom panel of Figure~\ref{fig:gammahist}
shows that, although there is significant overlap between the two distributions
at values of $\gamma \lsim 2$, most LSBs cluster around $\gamma \simeq 1.2$
compared to $\gamma \simeq 0.6$ for the halos.  We note, however, that the
contribution of a baryonic component has not been taken into account in our
analysis of the simulated halo $V_c$ profiles.  In order to investigate the
effect of a baryonic disk on the shape of the $V_c$ profile, we construct an
analytic mass model comprised of an NFW halo and an exponential disk.  We use
the prescription of \cite{MO98} to determine the scale length of the disk,
$R_d$, as a function of the concentration, $c$, and spin parameter, $\lambda$,
of the NFW halo, and the mass, $m_d$, and angular momentum, $j_d$, of the disk
(expressed as fractions of the halo mass and angular momentum, respectively).
Fitting eq.~\ref{eq:multifit} to the inner 20 kpc of the resulting $V_c$
profiles, we find that the best fit $\gamma$ value changes from $\gamma=0.72$
for an NFW halo with $c=10$ and $r_{200}=200~\kpch$, to $\gamma =0.92~(1.26)$
with the addition of a disk with $m_d=j_d=0.05~(0.1)$ assuming $\lambda=0.1$, as
expected for LSB disks.

We therefore conclude that the presence of the disk might be responsible for a
significant sift in the peak of the $\gamma$ distribution.

This suggests that the shape of most LSB rotation curves might actually be
consistent with the circular velocity profiles of $\Lambda$CDM halos.  The major
difference in the two distributions comes from considering the outliers; in
particular, systems with $\gamma \gsim 2.5$. Almost $1$ in $3$ LSBs is an
outlier according to this definition, but such high values of $\gamma$ are rare
amongst halos: fewer than $1$ in $40$ halos have $\gamma>2.5$, and fewer than
$1$ in $100$ have $\gamma>3$.  The ``cusp vs. core'' discrepancy alluded to
above appears confined to less than a third of LSBs; those with sharp
($\gamma\gsim 2.5$) turnovers in their rotation curves.

\begin{figure}
\plotone{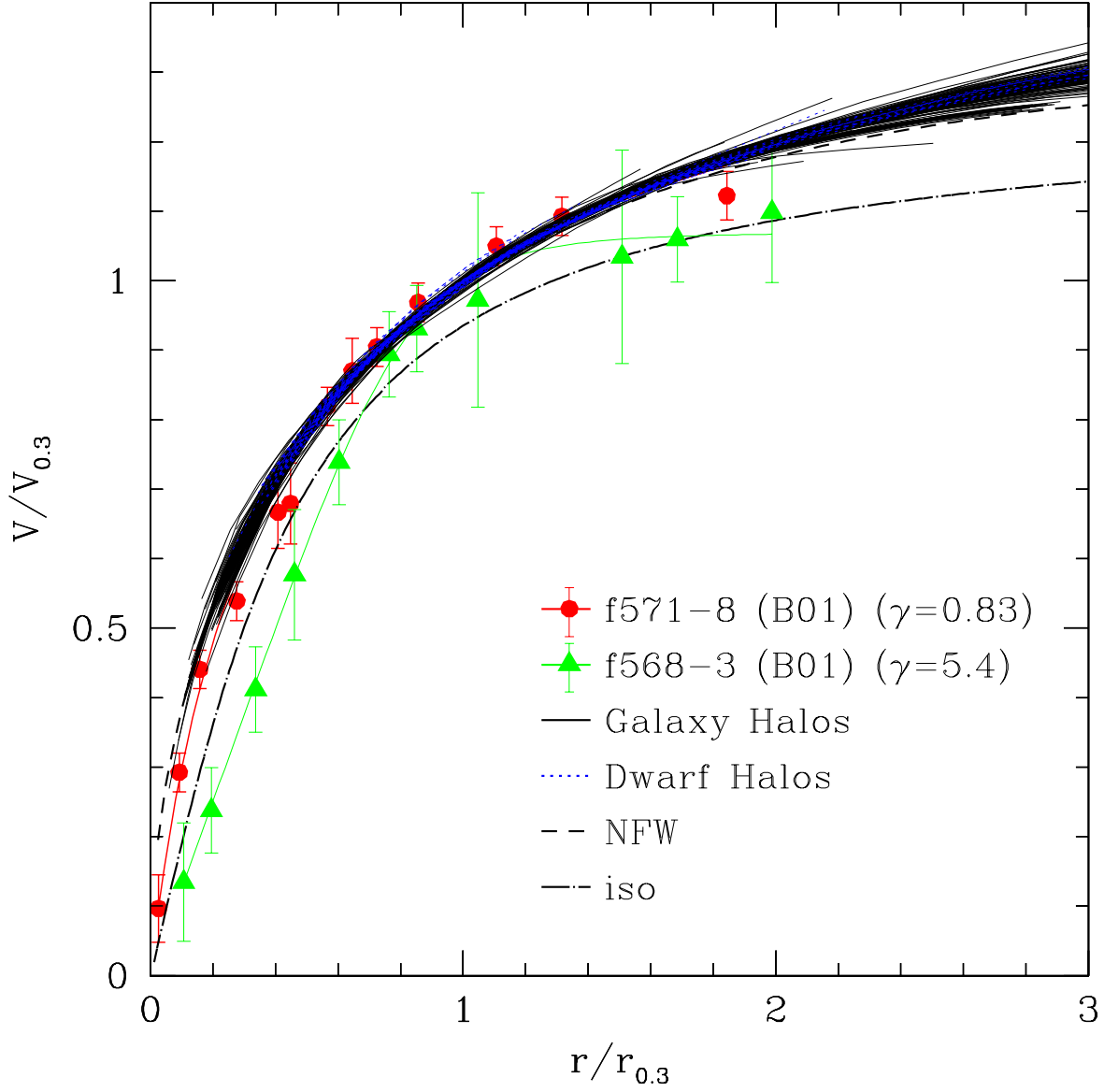}
\figcaption{
Rotation curves of two galaxies, one with $\gamma=0.83$ (f571-8, filled
circles), representative of the bulk of the population, and another one with
$\gamma=5.4$ (f568-3, filled triangles) representative of the tail of the
distribution. The curves have been scaled to the radius, $r_{0.3}$, (and
corresponding velocity, $V_{0.3}$) where the logarithmic slope of the fit to the
rotation curve is $d\log V/d\log r =0.3$. Galaxy (dwarf) halos are shown by solid
(dotted) lines, and they appear to follow closely the NFW profile (dashed
line). This figure shows clearly that halo $V_c$ profiles are inconsistent with
rotation curves with high $\gamma$, such as f568-3. The dot-dashed curve shows
that f568-3 is somewhat better matched by a non-singular isothermal sphere
model.\label{fig:multifit2}}
\end{figure}

We highlight the disagreement in Figure~\ref{fig:multifit2}, where we show the
$V_c$ profiles of all halos at $z \leq 1$ alongside two rotation curves; one
with $\gamma\sim 1$ and another with a relatively high $\gamma\sim 5$. In order
to concentrate on the {\it shape} of the rotation curve rather than on the
physical scaling parameters, profiles have been scaled in this plot to the
radius where the logarithmic slope of the fit to the rotation curve is equal to
$d\log V/d\log r =0.3$.  We shall refer to this radius and its
corresponding velocity as $r_{0.3}$ and $V_{0.3}$, respectively.  These
parameters are easily retrieved from the fits with eq.~\ref{eq:multifit}, and
are given by: $r_{0.3} = (7/3)^{-1/\gamma} \, r_t$ and $V_{0.3} =
(10/7)^{-1/\gamma} \, V_0$.

The scaled halo profiles are described reasonably well, on average, by the NFW
profile, shown as the dashed curve in Figure~\ref{fig:multifit2}, as is that of
F571-8 ($\gamma=0.83$). The shape of the rotation curve of F568-3 ($\gamma =
5.4$), on the other hand, is clearly inconsistent with the halo profiles. Better
(albeit not perfect) fits to galaxies with $\gamma\gsim 2.5$ may be obtained
using the circular velocity curve of a system with a constant density core, as
shown by the pseudo-isothermal{\footnote{The density profile of this widely used
approximation to the non-singular isothermal sphere is given by $\rho_{\rm
iso}(r)=\rho_0/(1 + (r/r_c)^2$, where $r_c$ is the core radius and $\rho_0$ is
the characteristic density of the core.}} model indicated by the dot-dashed
curve in Figure~\ref{fig:multifit2}.

\subsection{The concentration of LSB halos}
\label{ssec:conc}

The discussion of the preceding section focused on the {\it shape} of the
rotation curves and halo $V_c$ profiles. We now turn our attention to the
physical parameters of the fits, in order to address claims that LSB galaxies
are surrounded by halos of much lower concentration than expected in the
$\Lambda$CDM scenario \citep{MCGAUGH98,DEBLOK01}. We emphasize again that it is
important to characterize both the observational data and the simulations in a
way that is as independent as possible from fitting formulae or
extrapolation. \cite{ALAM02} recently proposed a simple and useful dimensionless
measure of mass concentration that satisfies these criteria,
\begin{equation}
\Delta_{V/2}  \equiv  \frac{\overline{\rho}(r_{V/2})}{\rho_{\rm crit}}.
\label{eq:deltav2}
\end {equation}

$\Delta_{V/2}$ measures the mean density contrast (relative to the critical
density for closure) within the radius at which the rotation speed drops to one
half of its maximum value, $V_{\rm max}$. In practice, we estimate ${\bar
\rho}(r)$ by $3\, V_c^2(r)/4\pi Gr^2$, a quantity that is easily measured both
in galaxies with rotation curve data (and well defined $V_{\rm max}$) and in
simulated halos.

\begin{figure}
\plotone{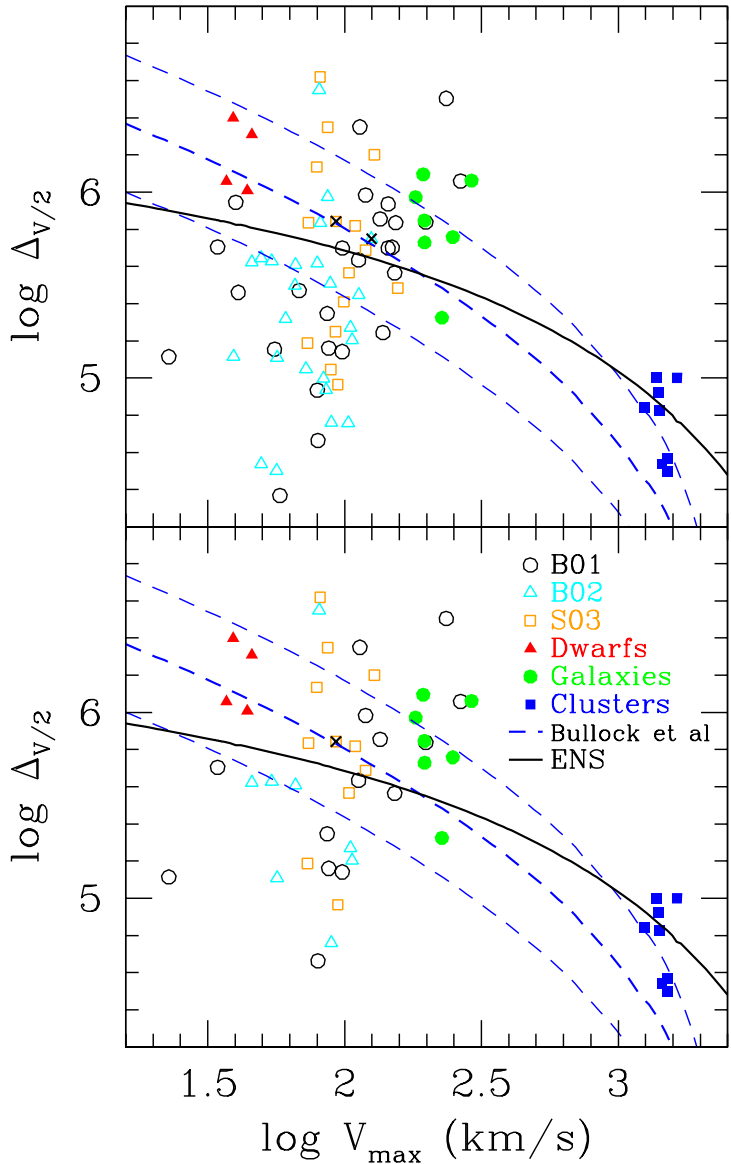}
\figcaption{
\cite{ALAM02} halo concentration parameter, $\Delta_{V/2}$, inferred from
rotation curves of LSB galaxies in the B01, B02, and S03 samples (open symbols),
compared to those measured for simulated dwarf- ($\log V_{\rm max} \simeq 1.6$),
galaxy- ($\log V_{\rm max} \simeq 2.3$), and cluster-sized ($\log V_{\rm max}
\simeq 3.2$) halos (solid symbols).  Points corresponding to $\Delta_{V/2}$
values for UGC 4325 calculated from B01 and S03 rotation curves are marked with
an $\times$.  Solid line shows the prediction of the \cite{ENS01} concentration
model for NFW halos in a \LCDM cosmogony ($\Omega_0=0.3$, $\Omega_\Lambda=0.7$,
and $h=0.65$, $\sigma_8= 0.9$).  Dashed lines show the prediction (and
$1~\sigma$ scatter) corresponding to the \cite{BULLOCK01} concentration model.
{\it Top panel:} All galaxies in combined B01, B02, and S03 dataset. {\it Bottom
panel:} Only galaxies whose rotation curves are not rising steeply at the
outermost point ($d\log V/d\log r (r_{\rm outer}) < 0.1$).
\label{fig:centdens}}
\end{figure}

The top panel of Figure~\ref{fig:centdens} shows $\Delta_{V/2}$ as a function of
$V_{\rm max}$ for all galaxies in the B01, B02, and S03 samples (open symbols),
together with the galaxy halos in our sample (filled circles). Filled triangles
and squares correspond to simulated dwarf galaxy and cluster halos, respectively
(see \S~\ref{sec:simulations}). 

The solid curve corresponds to the predictions of the \cite{ENS01} halo
concentration model for NFW halos in the \LCDM cosmology we have assumed for our
simulations.  The dashed curves show the predictions of the \cite{BULLOCK01}
concentration model, together with the 1-$\sigma$ halo-to-halo scatter predicted
by their model.  Figure~\ref{fig:centdens} shows that the simulations are in
rough agreement with both models; the \cite{BULLOCK01} model reproduces the
simulations slightly better on the scales of dwarf halos, whilst the ENS model
does better on the scale of clusters.

Figure~\ref{fig:centdens} also shows that, on average, the concentration of
$\Lambda$CDM is roughly comparable to those of LSBs in the samples we
considered, although the scatter appears much larger than either obtained in the
simulations or expected from the analytic model of \cite{BULLOCK01}. This result
is reminiscent of our findings regarding the distribution of rotation curve
parameter $\gamma$ (see Figure~\ref{fig:gammahist}): CDM halos appear consistent
with the {\it bulk} of LSBs but are at odds with the most deviant systems in the
sample.

In order to investigate the effect of observational uncertainties on this
parameter we examine the values of $\Delta_{V/2}$ calculated for UGC 4325.  As
discussed in \S\ref{sec:compobs}, the maximum velocities measured by B02 and
S03 differ by $30\%$ for this galaxy.  The corresponding values of
$\Delta_{V/2}$ are shown as the open triangle (B02) and open square (S03) points
in the top panel of Figure~\ref{fig:centdens}.  Surprisingly, we find that the
inferred halo central densities are very similar for the two versions of the
rotation curve.  Because $\Delta_{V/2}$ depends only on the ratio between
$V_{\rm max}$ and the radius $R_{V/2}$, two curves with very different maximum
velocities can give the same value of $\Delta_{V/2}$ provided that the rising
parts of the curves have the same slope.  The halo central densities one infers
from the B02 and S03 versions of the rotation curve for UGC 4325 are within
$20\%$ of one another despite a $30\%$ difference in $V_{\rm max}$ between the
two curves.  This level of uncertainty in $\Delta_{V/2}$ is negligible
considering that the halo-to-halo scatter in the simulations and model
predictions spans a factor of 6 at a given $V_{\rm max}$.  

Although the effect of observational uncertainties may be relatively unimportant
in the case of UGC 4325, we nonetheless attempt to reduce the scatter in the
$\Delta_{V/2}$ by culling from the sample all galaxies whose rotation curves are
still rising at the outermost radius probed by the data.  The bottom panel of
Figure~\ref{fig:centdens} shows the central densities inferred from rotation
curves with $d\log V/d\log r (r_{\rm outer}) < 0.1$.  We find that
this reduced sample retains much of the original scatter, and many points lie
both above and below the model predictions at a given value of $V_{\rm max}$.

Our results thus agree with those of \cite{ZENTNER02}, who have argued that the
large number of low-concentration galaxies in LSB samples calls for substantial
revision of the ``concordance'' $\Lambda$CDM scenario, such as tilted power
spectra, running spectral index, or perhaps a lower $\sigma_8$. However, it
appears unlikely that any such modification to the $\Lambda$CDM scenario would
result in the larger variety of rotation curve shapes and concentrations
apparently demanded by the LSB datasets. It is unclear at this point how to
reduce the disagreement, but any resolution to the puzzle must explain why so
many LSBs are actually in good agreement with $\Lambda$CDM halos and why the
disagreement is confined to a minority of systems. The possibility remains that
some complex astrophysical process not yet considered in the models might
actually be behind the discrepancy and that no radical modification to the
$\Lambda$CDM paradigm is called for.

\section{Conclusions}
\label{sec:concl}

We present results from a set of high resolution cosmological simulations of
dark matter halos formed in a \LCDM cosmogony.  Seven Milky Way-sized galaxy
halos were simulated at various mass, time, and spatial resolutions, enabling us
to investigate the convergence properties of cosmological N-body simulations. We
have examined the internal structure of the highest resolution realization of
each halo, with particular emphasis on the logarithmic slope of the inner
density profile. Finally, we have compared the circular velocity profiles of these
halos with the rotation curves of a large sample of dwarf and LSB galaxies.

Our main conclusions may be summarized as follows.

\begin{itemize}
\item 
The convergence criteria proposed by \cite{POWER03} are robust, and provide a
conservative estimate of the minimum radius at which the mass profile of
simulated halos can be reliably predicted.  According to these criteria, the
highest resolution galaxy halos we have simulated (which contain 2-4 million
particles within the virial radius) are reliably resolved down to $r_{\rm conv}
\simeq 1~\kpch$.

\item 
The slope of \LCDM halo density profiles becomes progressively shallow all the
way down to the minimum reliably resolved radius, without sign of converging to
a a well-defined power law near the centre. 

\item 
In general, the slope changes with radius more gradually than predicted by the
NFW formula, which leads some halos to be better described by profiles with
steeper cusps, such as the modification to the NFW formula proposed by Moore et
al (1999). There is, however, significant variation from halo to halo in the
radial dependence of the slope.  Some systems are better fit by the NFW profile,
and others by the Moore et al formula. At $r_{\rm conv}$, however, the density
profiles are significantly shallower than $r^{-1.5}$, the asymptotic value
advocated by \cite{MOORE99}.  

\item
A comparison of the circular velocity profiles of \LCDM halos with rotation
curves of dwarf and LSB galaxies indicates that the shapes of most ($\simeq
70\%$) LSB rotation curves are consistent with those of simulated halo $V_c$
profiles.  The remaining $\simeq 30\%$ feature a sharp turnover from the rising
to the flat part of the curve which is not consistent with the structure of
simulated halos.

\item 
The concentration of dwarf and galaxy-sized halos simulated in the
``concordance'' \LCDM cosmogony is in reasonable agreement with the
characteristic central density inferred for most LSB galaxies from rotation
curve data. However, as with the rotation curve shape, a significant number of
LSBs have concentrations well below (and above) the expected range.
\end{itemize}

We conclude that the inner structure of \LCDM halos is not manifestly
inconsistent with the rotation curves of LSB galaxies, although they seem unable
to reproduce the full variety of LSB rotation curve shapes and
normalizations. This discrepancy may signal the need to revise some of the basic
tenets of the \LCDM scenario, but it might also taken to imply that the relation
between gas rotation speeds and spherically-averaged halo circular velocities is
more complex than assumed in simple analyses such as the one presented here.

CDM halos, for example, are known to be triaxial, which may lead gaseous disks
to deviate systematically and significantly from simple coplanar circular
orbits. Work is in progress to try and determine whether such asymmetries in the
potential are able to account quantitatively for the ``cusp vs core''
discrepancy. Until this or other plausible astrophysical origin for the
discrepancy is identified, the observed variety of LSB rotation curves and
concentrations will remain a challenge that the theory must meet.

\acknowledgements

We thank Colin Leavitt-Brown for expert assistance with the IBM/SP3
supercomputer at the University of Victoria. We acknowledge useful conversations
with Stacy McGaugh, Stephane Courteau, David Hartwick, Chris Pritchet, Richard
Bower, Andrew Zentner, and James Bullock. EH is grateful for the hospitality of
the Max-Planck Institute for Astrophysics in Garching and the University of
Durham Department of Physics where some of this work was completed. The Natural
Sciences \& Engineering Research Council of Canada (NSERC) and the Canadian
Foundation for Innovation have supported this research through various grants to
JFN.

\bibliographystyle{astron}
\bibliography{stan}



\clearpage


\begin{deluxetable}{lrrrrrlr}
\tablecaption{Numerical and Physical Properties of Simulated Halos \label{tab:sims}}

\tablehead{
\colhead{Label}&\colhead{$R_{200}$}&\colhead{$M_{200}$}&\colhead{$N_{200}$}&\colhead{$\epsilon$}&
\colhead{$N_{\Delta t}$}& \colhead{Code} & \colhead{$r_{\rm conv}$} \\
&\colhead{$\kpch$}&\colhead{$\Msolh$}&&\colhead{$\kpch$}&&&\colhead{$\kpch$} }

\startdata

G1/$32^3$&202.1&191.8&5758&10&800&{\tt PKDGRAV}&22.9\\
G1/$64^3$&205.2&201.1&48318&2.5&1600&{\tt PKDGRAV}&7.7\\
G1/$128^3$&205.1&200.6&383560&1.25&3200&{\tt PKDGRAV}&3.2\\
G1/$256^3$&214.4&229.4&3447447&0.15625&{\tt RhoSgAcc}&{\tt GADGET}&1.4\\
&&&&&&&\\
G2/$32^3$&231.4&288.1&8583&10&800&{\tt PKDGRAV}&24.2\\
G2/$64^3$&231.8&289.8&69088&3.5&1600&{\tt PKDGRAV}&7.0\\
G2/$128^3$&234.1&298.6&566456&1.25&3200&{\tt PKDGRAV}&2.9\\
G2/$256^3$&232.6&292.9&4523986&0.5&6400&{\tt PKDGRAV}&1.3\\
&&&&&&&\\
G3/$32^3$&218.7&243.1&5484&10&800&{\tt PKDGRAV}&22.5\\
G3/$64^3$&215.0&231.1&41719&3.5&1600&{\tt PKDGRAV}&9.3\\
G3/$128^3$&214.5&229.7&331314&1.25&3200&{\tt PKDGRAV}&3.9\\
G3/$256^3$&212.7&223.8&2661091&0.45&6400&{\tt PKDGRAV}&1.7\\
&&&&&&&\\
G4/$64^3$&164.9&104.4&53331&2.5&3200&{\tt PKDGRAV}&5.7\\
G4/$128^3$&165.4&105.3&432313&1&6400&{\tt PKDGRAV}&2.4\\
G4/$256^3$&164.0&102.6&3456221&0.3&12800&{\tt PKDGRAV}&1.0\\
&&&&&&&\\
G5/$64^3$&165.3&105.0&62066&3&800&{\tt PKDGRAV}&5.7\\
G5/$128^3$&165.7&105.8&496720&1&3200&{\tt PKDGRAV}&2.3\\
G5/$256^3$&165.0&104.5&3913956&0.35&6400&{\tt PKDGRAV}&1.0\\
&&&&&&&\\
G6/$64^3$&160.7&96.5&57008&3&800&{\tt PKDGRAV}&6.0\\
G6/$128^3$&163.2&101.1&474844&1&3200&{\tt PKDGRAV}&2.5\\
G6/$256^3$&162.5&99.9&3739913&0.35&6400&{\tt PKDGRAV}&1.0\\
&&&&&&&\\
G7/$64^3$&159.7&94.7&55947&3&800&{\tt PKDGRAV}&5.8\\
G7/$128^3$&160.9&96.9&454936&1&3200&{\tt PKDGRAV}&2.4\\
G7/$256^3$&160.3&95.8&3585676&0.35&6400&{\tt PKDGRAV}&1.0\\
&&&&&&&\\
&&&&&&&\\
D1&32.3&0.8&784980&0.0625&{\tt EpsAcc}&{\tt GADGET}&0.3\\
D2&34.1&0.9&778097&0.0625&{\tt EpsAcc}&{\tt GADGET}&0.4\\
D3&32.3&0.8&946421&0.0625&{\tt EpsAcc}&{\tt GADGET}&0.3\\
D4&34.7&1.0&1002098&0.0625&{\tt EpsAcc}&{\tt GADGET}&0.3\\
&&&&&&&\\
&&&&&&&\\
C1&1502.1&78842.4&1565576&5.0&{\tt EpsAcc}&{\tt GADGET}&16.8\\
C2&1468.1&73618.2&1461017&5.0&{\tt EpsAcc}&{\tt GADGET}&16.9\\
C3&1300.6&51179.5&1011918&5.0&{\tt EpsAcc}&{\tt GADGET}&16.1\\
C4&1316.7&53101.9&1050402&5.0&{\tt EpsAcc}&{\tt GADGET}&15.9\\
C5&1375.5&60541.8&1199299&5.0&{\tt EpsAcc}&{\tt GADGET}&16.2\\
C6&1521.1&81870.6&1626161&5.0&{\tt EpsAcc}&{\tt GADGET}&15.5\\
C7&1245.8&44979.4&887837&5.0&{\tt EpsAcc}&{\tt GADGET}&16.4\\
C8&1365.4&59220.3&1172850&5.0&{\tt EpsAcc}&{\tt GADGET}&16.8\\

\enddata

\end{deluxetable}

\clearpage

\begin{deluxetable}{lrrrrrrl}
\tablecaption{Properties of Rotation Curves and Fit Parameters \label{tab:multifit}}
\tablehead{
\colhead{Galaxy ID} & \colhead{$r_{\rm max}$}&
\colhead {$V_{\rm max}$} &\colhead{$V(r_{\rm outer})$}&
 \colhead {$r_t$}&\colhead{$\gamma$}&\colhead{$V_0$}&
\colhead {$\chi^2_{\rm red}$}\\
& \colhead{(kpc)}&\colhead {(km/s)} &\colhead{(km/s)}&
 \colhead {(kpc)}&&\colhead{(km/s)}&}
\startdata
\cite{DEBLOK01}: \\
ESO0140040&21.2&272.7&262.8&3.5&1.5&272.7&$5.26 \times 10^{-1}$\\
ESO0840411&8.9&61.3&61.3&10.7&1.4&105.4&$2.22 \times 10^{-1}$\\
ESO1200211&3.5&25.4&25.4&1.1&5.0&22.8&$4.41 \times 10^{-2}$\\
ESO1870510&2.7&39.9&39.9&1.5&1.1&55.9&$2.49 \times 10^{-2}$\\
ESO2060140&11.7&118.0&118.0&1.9&1.6&123.9&$9.74 \times 10^{-2}$\\
ESO3020120&10.0&86.3&86.0&3.3&2.6&88.0&$2.51 \times 10^{-3}$\\
ESO3050090&4.4&54.6&54.0&3.1&1.6&71.3&$5.23 \times 10^{-2}$\\
ESO4250180&14.4&144.5&144.5&5.0&1.1&177.2&$9.18 \times 10^{-2}$\\
ESO4880049&6.0&97.1&97.1&2.5&1.1&128.5&$6.20 \times 10^{-3}$\\
F563-1&13.4&112.4&110.9&2.2&1.1&121.8&$9.97 \times 10^{-2}$\\
F568-3&11.2&101.1&101.1&4.8&5.4&98.4&$3.14 \times 10^{-2}$\\
F571-8&14.0&143.9&143.9&2.8&0.8&196.7&$6.22 \times 10^{-1}$\\
F579-v1&11.6&114.4&114.2&1.1&1.4&115.7&$3.73 \times 10^{-2}$\\
F583-1&14.0&86.9&86.9&4.0&2.0&91.1&$9.79 \times 10^{-3}$\\
F583-4&6.7&69.9&69.9&1.2&0.6&122.2&$1.67 \times 10^{-1}$\\
F730-v1&11.9&145.3&145.3&2.2&1.3&157.3&$7.43 \times 10^{-2}$\\
UGC11454&11.9&152.2&152.2&2.8&1.3&172.1&$4.36 \times 10^{-1}$\\
UGC11557&6.2&80.4&80.4&14.1&1.0&264.3&$4.61 \times 10^{-2}$\\
UGC11583&1.5&35.6&35.6&0.9&4.7&35.0&$3.28 \times 10^{-2}$\\
UGC11616&9.6&142.8&142.8&2.4&1.7&142.2&$1.98 \times 10^{-1}$\\
UGC11648&12.7&144.6&144.6&57.5&-0.2&13.0&$4.72 \times 10^{-1}$\\
UGC11748&5.3&250.0&246.5&2.3&10.4&235.4&$1.95 \times 10^{0}$\\
UGC11819&8.9&154.7&152.6&4.6&3.3&154.7&$4.22 \times 10^{-1}$\\
UGC4115&1.0&39.8&39.8&2.1&1.2&115.9&$2.20 \times 10^{-3}$\\
UGC5750&21.8&78.9&78.9&7.0&3.4&80.5&$8.95 \times 10^{-3}$\\
UGC6614&45.4&205.2&203.9&2.8&1.2&201.9&$2.30 \times 10^{0}$\\
&&&&&&&\\
\cite{DEBLOK02}: \\
DDO185&2.2&49.6&49.6&2.3&1.1&76.2&$1.88 \times 10^{0}$\\
DDO189&6.4&65.7&64.4&1.7&1.7&68.8&$5.33 \times 10^{-2}$\\
DDO47&3.2&67.0&67.0&2.9&1.7&88.0&$2.75 \times 10^{-1}$\\
DDO52&3.1&50.0&50.0&1.2&79.4&45.9&$3.34 \times 10^{0}$\\
DDO64&2.7&46.9&46.9&1.6&2.0&58.0&$5.06 \times 10^{-1}$\\
IC2233&7.4&92.8&92.8&5.1&1.1&135.2&$1.39 \times 10^{0}$\\
F563-1&17.5&114.1&114.1&2.8&1.2&123.9&$2.67 \times 10^{-1}$\\
NGC100&8.0&91.3&91.3&2.6&1.8&95.0&$8.90 \times 10^{-2}$\\
NGC1560&7.9&77.5&77.5&1.8&1.0&94.8&$8.30 \times 10^{0}$\\
NGC2366&1.4&55.5&54.6&1.5&20.1&54.2&$2.41 \times 10^{-1}$\\
NGC3274&2.1&82.6&79.5&0.7&1.9&81.2&$7.82 \times 10^{-1}$\\
NGC4395&7.6&84.2&82.6&1.2&1.0&92.8&$5.65 \times 10^{-1}$\\
NGC4455&5.9&64.4&64.4&1.8&0.9&92.7&$1.96 \times 10^{-1}$\\
NGC5023&5.9&84.4&84.4&1.4&1.4&95.0&$2.57 \times 10^{-1}$\\
UGC10310&9.0&75.0&75.0&3.7&1.6&82.7&$2.76 \times 10^{-1}$\\
UGC1230&14.1&112.7&102.9&4.3&10.6&105.1&$6.43 \times 10^{-2}$\\
UGC1281&5.2&56.9&56.9&3.0&5.4&57.2&$7.14 \times 10^{-3}$\\
UGC3137&13.5&106.9&104.6&4.8&3.3&106.6&$8.77 \times 10^{-2}$\\
UGC3371&10.3&85.7&85.7&5.6&1.7&101.9&$2.76 \times 10^{-3}$\\
UGC4173&12.2&57.0&57.0&5.5&0.6&119.2&$9.59 \times 10^{-3}$\\
UGC4325&4.6&122.6&122.6&4.6&1.4&208.7&$1.37 \times 10^{-2}$\\
UGC5005&22.0&100.0&99.1&6.6&1.2&119.9&$4.75 \times 10^{-2}$\\
UGC5750&5.7&49.6&49.6&6.4&3.8&63.7&$1.27 \times 10^{-2}$\\
UGC711&15.4&91.6&91.6&6.8&3.1&91.8&$6.55 \times 10^{-2}$\\
&&&&&&&\\
\cite{SWATERS03}: \\
F563V2&7.5&113.1&111.2&2.5&7.0&109.4&$3.10 \times 10^{-1}$\\
F5681&12.4&130.7&130.7&1.4&1.3&133.6&$3.49 \times 10^{0}$\\
F5683&5.6&111.2&100.6&5.1&52.2&96.6&$1.85 \times 10^{0}$\\
F568V1&8.9&124.9&118.2&2.5&1.7&121.9&$7.99 \times 10^{-1}$\\
F5741&11.5&104.2&102.6&2.4&1.3&112.2&$5.58 \times 10^{-1}$\\
UGC11557&10.4&84.5&84.5&6.0&0.7&189.7&$6.27 \times 10^{-1}$\\
UGC11707&15.0&99.9&99.9&1.5&0.6&149.6&$3.94 \times 10^{-1}$\\
UGC11861&8.1&164.0&152.6&3.3&0.9&200.0&$4.61 \times 10^{0}$\\
UGC2732&15.4&98.0&98.0&1.7&0.7&127.6&$6.89 \times 10^{-1}$\\
UGC2259&1.7&93.7&90.0&0.7&1.5&88.5&$1.83 \times 10^{0}$\\
UGC4325&2.8&104.6&91.5&1.8&3.7&93.5&$8.31 \times 10^{-1}$\\
UGC4499&8.5&74.3&74.3&3.4&3.7&73.9&$1.13 \times 10^{0}$\\
UGC5721&2.2&80.4&78.7&0.4&1.0&86.0&$7.28 \times 10^{-1}$\\
UGC731&5.8&74.0&73.9&1.3&1.4&78.7&$9.74 \times 10^{-1}$\\
UGC8490&6.1&80.1&77.6&1.2&2.7&79.4&$5.56 \times 10^{-1}$\\

\enddata


\end{deluxetable}
\onecolumn
\clearpage

\clearpage

\clearpage

\clearpage

\clearpage

\clearpage

\clearpage

\clearpage

\clearpage

\clearpage

\clearpage

\end{document}